\documentclass[12pt,preprint]{aastex}

\begin{document}
\title{Birth Kick Distributions and the Spin-Kick Correlation of Young Pulsars}
\author{C.-Y. Ng\altaffilmark{1} \& Roger W. Romani\altaffilmark{1}}
\altaffiltext{1}{Department of Physics, Stanford University, Stanford, CA 94305}
\email{ncy@astro.stanford.edu, rwr@astro.stanford.edu}

\begin{abstract}
	Evidence from pulsar wind nebula symmetry axes and radio polarization
observations suggests that pulsar motions correlate with the spin directions.
We assemble this evidence for young isolated pulsars and show how it can be used
to quantitatively constrain birth kick scenarios. We illustrate by computing
several plausible, but idealized, models where the momentum thrust is 
proportional to the neutrino cooling luminosity of the proto-neutron star.
Our kick simulations include the effects of pulsar acceleration and spin-up
and our maximum likelihood comparison with the data constrains the model parameters.
The fit
to the pulsar spin and velocity measurements suggests that: $i$) the anisotropic
momentum required amounts to $\sim10$\% of the neutrino flux, $ii$) while a
pre-kick spin of the star is required, the preferred magnitude is small
$10-20\;\mathrm{rad\;s^{-1}}$, so that for the best-fit models $iii$) the bulk of
the spin is kick-induced with $\bar \Omega \sim 120\;\mathrm{rad\;s^{-1}}$
and $iv$) the models suggest that the anisotropy emerges on a timescale $\tau \sim 1-3$s.
\end{abstract}
                                                                                

\section{Introduction} 

	The high velocity of pulsars, at least an order of magnitude
larger than their parent population \citep{hob05}, has been a long standing
puzzle. Of the many models proposed, all except the classic \citet{bla61}
binary break-up mechanism rely directly or indirectly on the large
release of gravitational binding energy associated with the core collapse
event. Since the fastest observed pulsars have $v > 10^3\;\mathrm{km\;s^{-1}}$
\citep{cor93,cha05}, i.e. bulk velocities $> 3 \times 10^{-4}\;$c, binary break-up
is clearly inadequate and a viable mechanism must tap an appreciable fraction of
the collapse momentum. However, a wide range of physical processes can still
be invoked \citep*[e.g.][]{lai01,jan05}, so gross energetic constraints alone do
not solve the pulsar kick problem.

	Recent progress has come from studies that probe the {\it vector}
nature of the kick. For example \citet{fry06} argue that neutrino-driven kicks
(which may even be crucial for successful explosion) will produce fast supernova
remnant (SNR) velocities along the pulsar velocity vector, while ejecta-driven
(recoil) kicks should have fast ejecta opposite to the pulsar motion. The pulsar
itself also contains kinematic evidence of the kick geometry, since both
the velocity and spin directions are in principle measurable. These linear and
angular momenta are fossils of the core collapse which can remain inviolate
for as much as $10^6\;$yr before acceleration in the Galactic potential perturbs
the velocity or torques perturb the spin. Thus mechanisms that are spin-dependent
can leave a fossil record in the kick distribution \citep{lai01,rom05}.
Conversely, kicks can leave a record in the spin distribution \citep{spr98}.

	The key to making such tests is accurate measurement of pulsar
motions and pulsar spins. For the former, the most important progress
has been the dramatic improvement in the quality and quantity of
VLBI astrometric proper motions \citep{cha05}. This work makes
it possible to have a vector proper motion (and in a number of cases a
parallax, and hence a 2-D space velocity) for most moderately bright,
$\geq$ few mJy, pulsars within 2-3\,kpc. An even larger sample of approximate
proper motions has been obtained from refined analysis of long term
timing records \citep{hob05}. Together these methods provide useful
space velocities for several dozen young pulsars within 3~kpc. Association
with parent supernova remnants can provide a few additional proper motion
estimates for interesting pulsars at larger distances.

More difficult to measure is the spin direction. The classic approach is to adopt
the `rotating vector' (RV) model for pulsar polarization and assign the projected 
position angle (PA) of the spin axis to the absolute (Faraday `rotation measure'
RM corrected) position angle of the polarization at the phase of maximum
PA sweep. This method suffers major systematic uncertainties due to poor matches
to the expected RV model sweep as well as an intrinsic $\pi/2$ ambiguity according
to whether the emission is in the normal or orthogonal mode; early analyses
\citep{des99} were not at all conclusive. However, recent improved polarization
measurements and RM corrections have provided convincing statistical evidence for
a projected 2-D correlation between the directions of polarization PA and the
proper motion \citep*{joh05,wan06a}. Even better, in several cases excellent
{\it CXO} images have made it possible to measure accurately the pulsar spin
position angle and its inclination from the plane of the sky using the symmetry
axis of the surrounding pulsar wind nebula (PWN) \citep{ng04}. Finally, in a few
cases, one can also independently estimate the original post-supernova spin of the
pulsar. With the 3-D spin orientation and amplitude, comparison with the projected
2-D motion allows even more powerful tests of the nature of the birth kick.

	In this paper we explore the possibility of comparing the new vector data on
pulsar spins and speeds with the dynamics of core collapse. The structure of the
paper is as follows. In section \S\ref{s21}, we describe three simple models of
the luminosity and size evolution of a proto-neutron star, which we will follow
in estimating the effects of a birth kick. In section \S\ref{s22}, we summarize
the effect of momentum deposition on kick and spin \citep[following][]{spr98}
and show how the integrated kick controls the birth properties of the pulsar. 
We also comment on the results of \citet*{wan06b}, whose related but somewhat
simpler study of kick evolution appeared while we were writing up this paper.
In Section \S\ref{s3}, we discuss the sample of nearby and/or young pulsars that
we will use for comparison with the models, emphasizing the important objects for
which we have more complete knowledge of the post-kick state of the star.
Section \S\ref{s4} describes a population simulation that allows us to
compare the {\it distribution} of birth properties with the observed neutron stars
for the several models. Section \S\ref{s5} summarizes the fit results that
constrain the basic parameters in the kick models. We conclude by discussing the
status of the spin-kick comparison, the insight that the present data give us
into the physics of the pulsar kick, and the prospects for further refinement of
these tests.

\section{Core-Collapse Luminosity-Driven Kicks} 

In this study, we consider some simple models for a kick associated with
the large luminosity of the post-bounce proto-neutron star. Our assumptions
are: $i)$ that the asymmetric momentum (kick) scales roughly as the emergent
neutrino luminosity, possibly with a delayed onset; $ii)$ that the {\it net} thrust
originates from a region fixed with respect to the star; and $iii)$ that the
thrust may be at a finite angle to the local normal. Of course, this net thrust
may simply be the surface average of a complex distribution of local forces.
However, we assume that the sum of these has a stable component with respect to
the star on the timescale of the initial rotation and cooling.  If the thrusts are 
impulsive and random then, as argued by \citet{spr98} and illustrated numerically
by \citet{wan06b}, the final velocity and spin tend to become randomly oriented.
Note also, as discussed by \citet{spr98}, that if there is no initial rotation, 
the result of a single kick is an orthogonal spin; we will show that the preferred
models have a small initial rotation, but the bulk of the spin is introduced by 
the kick itself.

	The simplest realization of this picture would be asymmetric neutrino 
emission mediated by strong $B-$fields. \citet{arr99a,arr99b} suggest that due to weak
interaction parity non-conservation, the neutrino emission depends on the magnetic
field, both through simple field-dependent opacities and through helicity-sensitive
scattering in the presence of a strong field. They suggest that the total asymmetry
in the neutrino flux is $\sim 0.1B_{15} \bar{E}_\nu^{-2}+0.002B_{15}/T$ where
$B=10^{15}B_{15}\,$G is the magnetic field strength, $\bar{E}_\nu$ is the mean
neutrino energy in MeV and $T$ is the temperature in MeV. Clearly, very large fields
are needed to produce a significant net kick. However, this model does provide a
natural `anchor point' on the star and, through the field angle at the surface, a
plausible mechanism for non-normal kicks.

	Other possibilities also exist. For example, the net momentum might be generated
by asymmetric neutrino coupling to matter in the envelope \citep{soc05}, or by `neutrino
star-spots' induced by the magnetic field \citep{tho93}, or even by hydrodynamic
motions associated with accretion onto the neutron star \citep[e.g.][]{blo06}. In the
last case, fall-back instabilities may apparently build to large amplitude. \citet{sch06}
have presented a detailed set of 2-D hydrodynamic simulations suggesting that large
recoil velocities of neutron stars occur naturally when $m=1$ instabilities with strong
narrow down-flows develop. Here the kick velocity is dominated by the gravitational
attraction of the dense down-flow. (A neutrino hot spot associated with this flow does in
fact generate anisotropic emission, but its kick, a $\sim 10$\% correction, is opposite
that of the gravitational term). Our model does not, of course, capture the full
dynamical content of such hydrodynamical kicks, and certainly the neutrino
light curve does not fully describe the thrust. However
this luminosity profile does provide a characteristic duration
for energetic processes associated with the core collapse. The physical interpretation is
thus not as clear as for the direct neutrino kick scenario, but the characteristic energetics
and time scale modeled in our fits can still probe the underlying physics.

\subsection{Proto-Neutron Star Luminosity Models\label{s21}}

\begin{figure}
\plotone{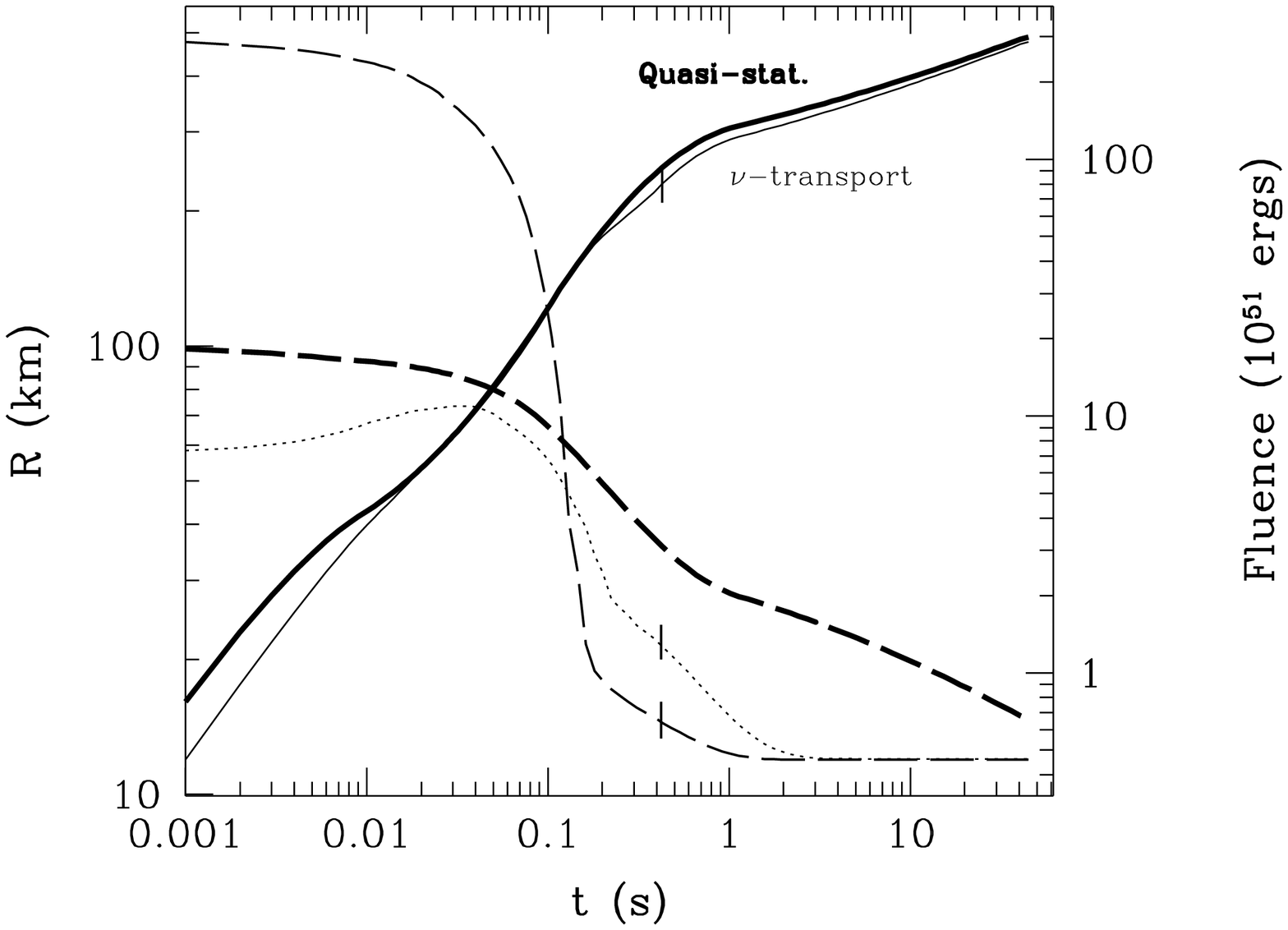}
\caption{Proto-neutron star cooling. Evolution of the stellar radius and the neutrino
fluence for the quasi-stationary model are shown in the thick solid and dashed lines
respectively. The thin solid and dashed lines show equivalent values for the 
$\nu$-transport model \citep{jan03}, with the effective $R$ of a uniform density
sphere with the same instantaneous moment of inertia. The thin dotted line shows
the radius of the neutrino-sphere in the $\nu$-transport model. The end points of
the numerical simulation are marked; the curves are extrapolated to $t=50\;$s.
\label{f1}}
\end{figure}

We first start from a simple cooling model, where the neutron star is optically
thick to neutrinos and the hot envelope is a dominant contribution to the moment
of inertia. If the star's gravitational binding energy is promptly released and
lost as neutrino radiation from the surface of the proto-neutron star, then
the proto-neutron star is in thermal equilibrium and the evolution of its radius
$R_{NS}$ is {\it `quasi-stationary'} and determined by the conservation of energy:
\[ L_\nu(t) \sim \frac{d}{dt}\frac{GM^2}{R_{NS}} \ . \] We show here results using
the neutrino luminosity $L_\nu$ from Fig.~3 in \citet{bur92} but have tried
different $L_\nu(t)$ from the literature. The results are quite insensitive to the
details of any standard cooling curve. In the simulation, the initial proto-neutron
star radius is set to a generic value of 100\,km. As shown in Fig.~\ref{f1}, a final
radius of 14.7\,km is obtained, which is determined by the total neutrino fluence.

	At the opposite extreme, we can assume that both the envelope and core are
highly degenerate at bounce \citep[for reviews:][]{jan04,kot06}. In this case,
$R_{NS}$ is independent of $L_\nu(t)$ and the kick essentially acts at a constant
radius throughout the cooling. We call this extreme the {\it `static'} model.
In our simulations we take this as $R_{NS}$=10\,\,km, constant throughout the kick.

\begin{figure}
\includegraphics[scale=0.8]{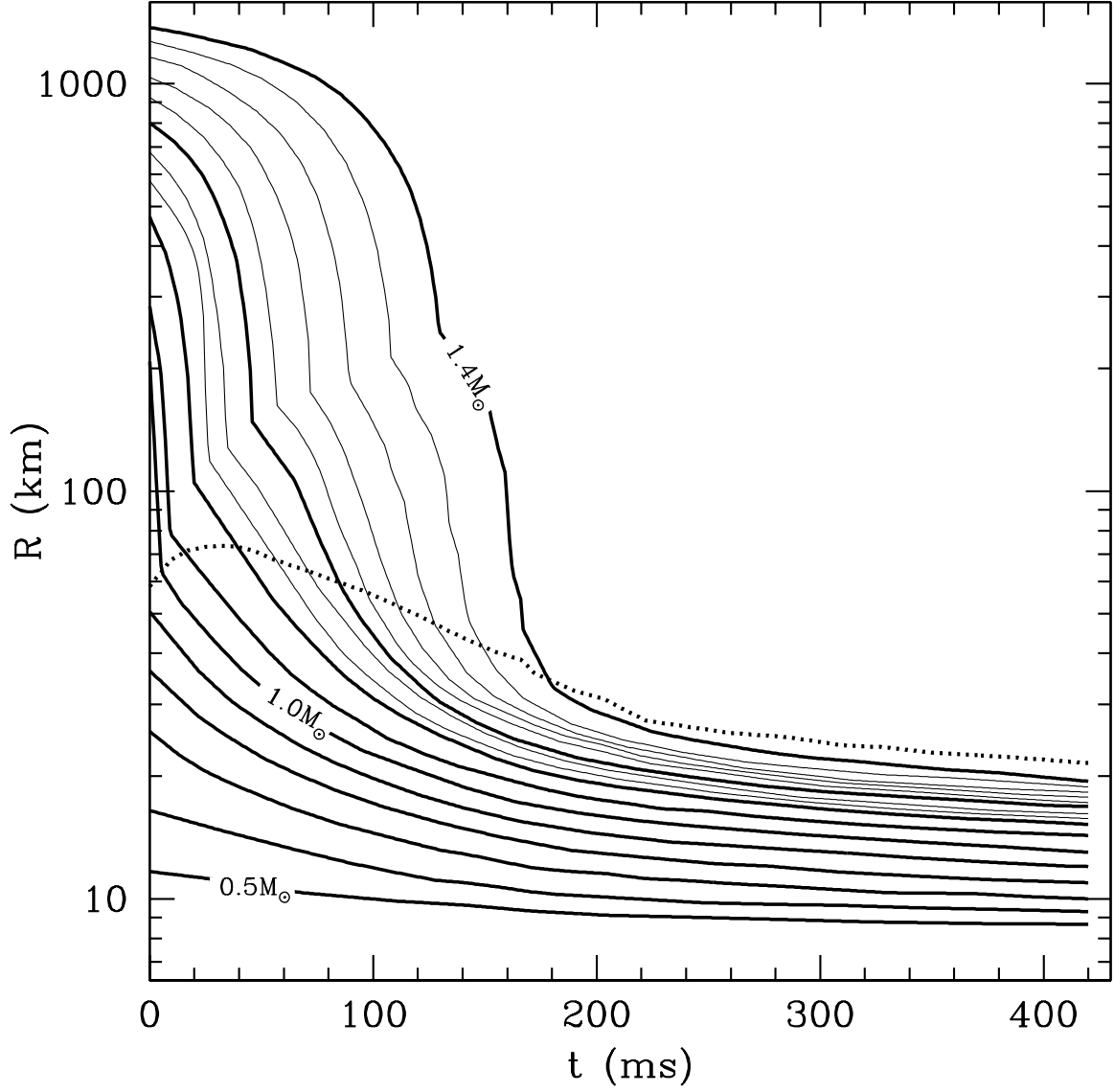}
\caption{Mass shell trajectories of a proto-neutron star in the $\nu$-transport
model, reproduced from \citet{jan03}. Radius of the neutrinosphere is plotted by
the dotted line.
\label{f2}}
\end{figure}

Results from more physical numerical simulations suggest real neutron star
behavior is between the two limiting cases above. To illustrate this, we examined
the results of 2-D hydrodynamic simulations from \citet{ram02} and \citet{jan04}.
These authors employ a Multi-Dimensional Boltzmann Transport and Hydrodynamics (MDBTH)
code to follow the neutrino advection and diffusion. In Fig.~\ref{f2} we reproduce the
mass shell trajectories and show the $\nu_e$ photosphere as it evolves during the
initial cooling. As one sees, the neutron star initially has a large but optically
thin envelope. We will assume the thrust is imparted near the neutrinosphere, but the
whole star is forced to co-rotate. The time varying moment of inertia can be followed
by integrating across these shells. The simulations using this evolution are referred
to as  {\it `$\nu$-transport'} models.  Fig.~\ref{f1} compares the basic behavior
for the models showing $L_\nu(t)$ and the effective $R_{NS}(t)$. The  {\it `static'}
model has, by assumption the same neutrino light curve as the {\it `quasi-stationary'}
picture, albeit at fixed $R_{NS}$. In the case of the {\it `$\nu$-transport'} model,
the simulations available stopped at $t=0.46\;$s (marked). We extrapolate with simple
matched exponentials to take the curves to 50\,\,s.

\subsection{Kick Dynamics\label{s22}}

\begin{figure}
\plotone{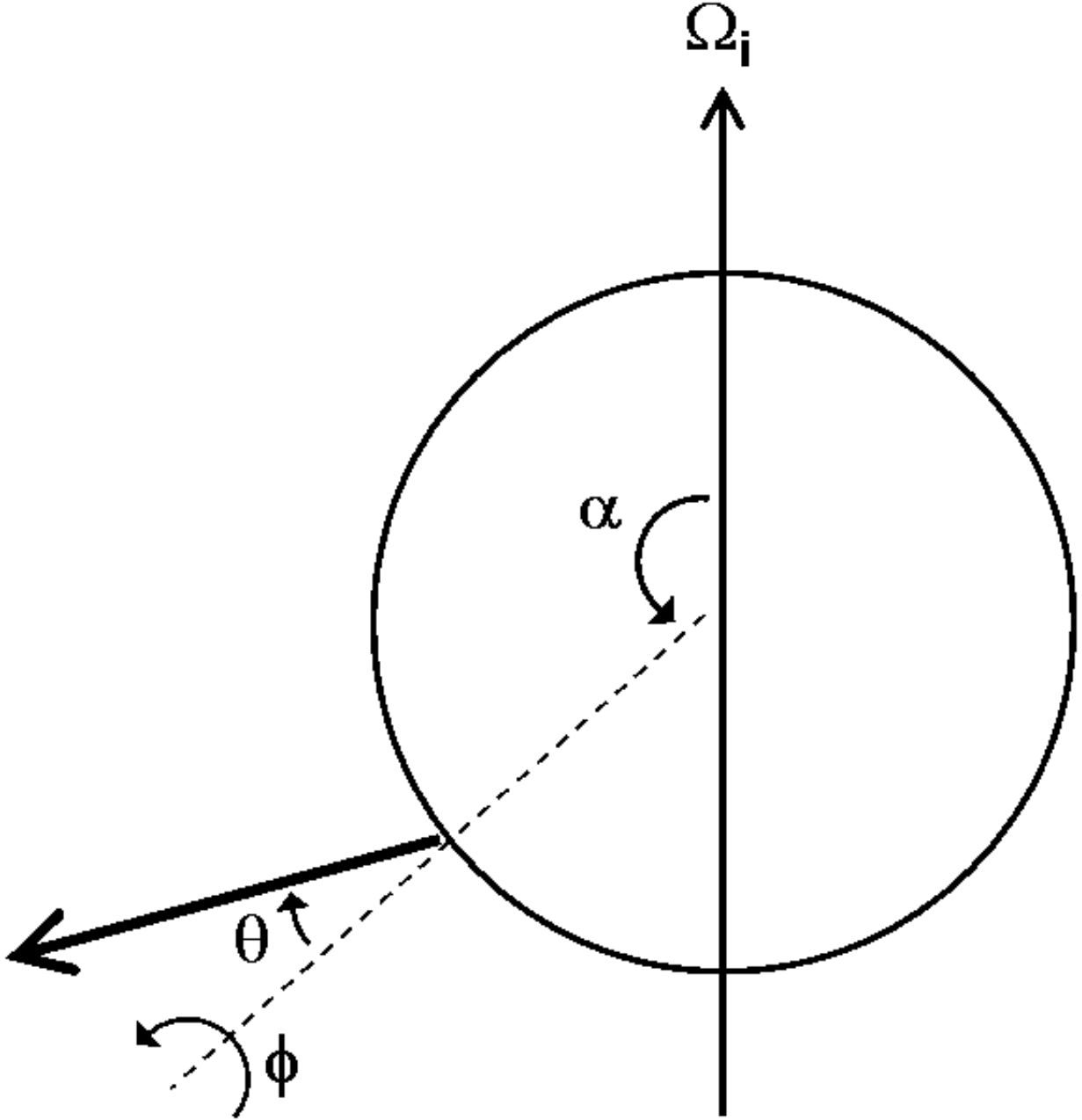}
\caption{The kick geometric parameters: $\alpha$ is the polar angle of the kick position,
$\theta$ is the off-axis angle of the kick from the surface normal, $\phi$ is the
azimuth angle of the kick about the normal, $\Omega_i$ is the pre-kick angular velocity.
\label{f3}}
\end{figure}

	We wish to model a (possibly non-normal) net thrust at
a fixed point $\vec{r}$ on the proto-neutron star surface which scales with
the $L_\nu(t)$ multiplied by a constant asymmetry $\eta$.
The thrust is given by \[ F_\nu(t)\sim \eta\frac{L_\nu}{c}\ . \]
The angular momentum $\vec{J}=I\vec{\Omega}$ evolves according to 
\[\vec{J} = \int \vec{r}\times \vec{F}_\nu \; dt + I_i\vec{\Omega}_i \ , \]
where $I_i$ and $\vec{\Omega_i}$ are the initial moment of inertia and angular velocity
before the kick respectively. Similarly, the linear momentum is determined by
\[ \vec{p} = \int \vec{F}_\nu\; dt \ . \]
As we are modeling isolated young neutron stars, three angles determine the kick
geometry as illustrated in Fig.~\ref{f3}: $\alpha$ is the polar angle of the kick
position, $\theta$ is the off-axis angle of the kick from the surface normal, and
$\phi$ is the azimuth angle of the kick about the normal, with $\phi=0$ closest to the initial
spin axis. Note that we are ignoring any initial space velocity of the progenitor,
since with typical values of only a few $\mathrm{km\;s^{-1}}$, it is negligible
compared to the kick velocity of several hundred $\mathrm{km\;s^{-1}}$. However, some
modest fraction of presently isolated young neutron stars are released from the
break-up of pre-existing binaries \citep{dew87,tau98,fry98}. In this case,
these may have an initial velocity of as much as 100\,$\mathrm{km\;s^{-1}}$ (the Blaauw mechanism).
If there is no relation between this initial velocity and the initial spin,
then this pre-natal velocity may be added to the post-kick speed in quadrature.
The main effect is to blur the correlations described below. If, however, there
is a fixed relation between the orbital speed and the pre-kick spin or the
kick itself, then this fourth angle must be introduced to the geometry. As these
cases represent a small fraction of the single neutron star population, we ignore 
them here; however such considerations may be important to descriptions of kicks for
neutron stars presently found in binaries. We defer discussion of such models
to a future communication.

During the kick, directions of $\vec{r}$ and $\vec{F}_\nu$
evolve according to the rotation
\[ \hat{r}\,^\prime = \mathcal{R} \hat{r} \] 
\[ \hat{F}_\nu^{\,\prime} = \mathcal{R} \hat{F}_\nu \ \] 
where the rotation matrix $\mathcal{R}$ is given by
\[ \mathcal{R} = \left( \begin{array}{ccc}
\Omega_x^2 + (1-\Omega_x^2)\cos\delta & \Omega_x\Omega_y(1-\cos\delta)-\Omega_z\sin\delta & \Omega_x\Omega_z(1-\cos\delta)-\Omega_y\sin\delta \\
\Omega_x\Omega_y(1-\cos\delta)-\Omega_z\sin\delta & \Omega_y^2 + (1-\Omega_y^2)\cos\delta & \Omega_y\Omega_z(1-\cos\delta)+\Omega_x\sin\delta \\
\Omega_x\Omega_z(1-\cos\delta)-\Omega_y\sin\delta &  \Omega_y\Omega_z(1-\cos\delta)+\Omega_x\sin\delta & \Omega_z^2 + (1-\Omega_z^2)\cos\delta
\end{array} \right) \ , \]
$\vec{\Omega}=\Omega\hat{\Omega}=\Omega(\Omega_x, \Omega_y, \Omega_z)$
is the angular velocity and $\delta=\Omega\Delta t$ is the angle that the
star rotates in time $\Delta t$.

Given a set of initial conditions, the above equations can be integrated numerically
to obtain the kick velocity, initial spin period and alignment angle.
Note that in this paper, we call the initial pulsar spin period $P_0$ which is the
value right after the kick. The pre-kick spin is denoted by $P_i$ ($\Omega_i$),
which is a specified parameter of the model.

\begin{figure}
\plotone{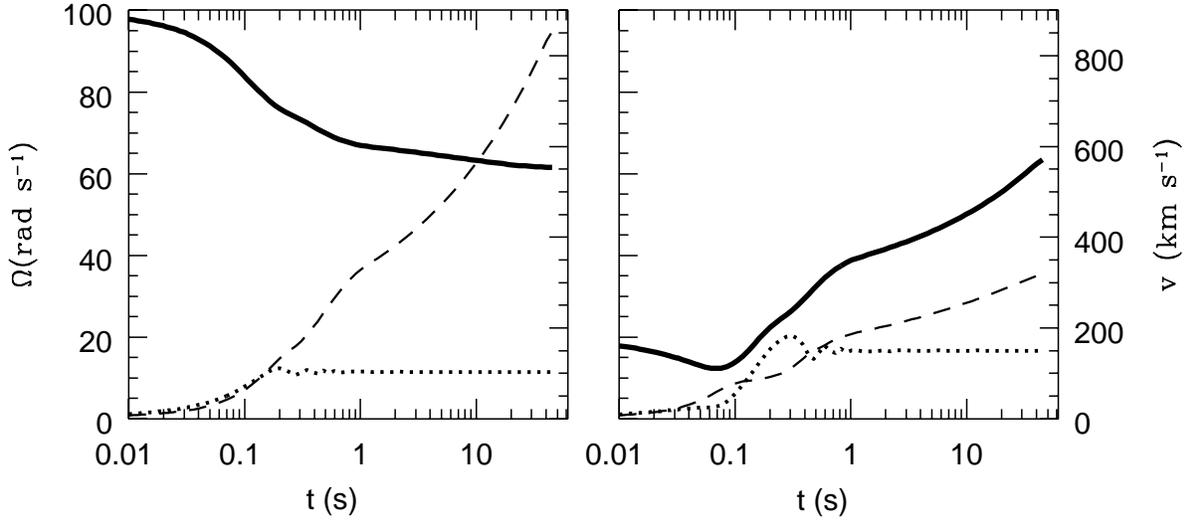}
\caption{Two kick integrations with identical parameters ($ \alpha=\cos ^{-1}0.6$, 
$\phi=80\arcdeg$, %i.e. at 10\arcdeg against the initial spin, 
$\theta=0.2\;$rad),
except
$\Omega_i=100\;\mathrm{rad\;s^{-1}}$ (\emph{Left panel}) and
$\Omega_i=30\;\mathrm{rad\;s^{-1}}$ (\emph{right panel}).
The solid lines follow the instantaneous angular momentum, plotting
the equivalent angular velocities with the moment of inertia 
at its final value.
The dashed and dotted lines are the parallel and perpendicular components of the 
space velocity with respect to the instantaneous spin direction.
\label{f4}}
\end{figure}

Fig.~\ref{f4} shows two example integrations with identical initial parameters except
for the pre-kick angular velocities $\Omega_i$. The kick is against the initial spin
with a component along the surface, thus the angular velocity decreases. However, with
$\Omega_i=30\;\mathrm{rad\;s^{-1}}$, the right panel shows that the 
torque from the kick is large enough to perturb the spin axis to a new direction,
where the kick is more effective in spinning up the proto-neutron star. Note that the
initially fast spinner ends up nearly aligned, while the slow spinner is
closer to orthogonal. Thus, even though the post-kick spin periods $P_0$ are
quite similar in the two models, information about the pre-kick spin is encoded
in the vector direction of ${\vec \Omega_0}$. In practice, the typical direction 
and degree of alignment also depends on the thrust orientation and duration; we show
below that comparison with multiple observables can separately probe these
various effects. 

\section{Pulsar Sample\label{s3}}

Linear and angular momenta are relics of neutron star birth and, if well preserved,
can provide important probes of the core collapse physics. Given their vector nature,
the alignment angle between them is an important observable. As discussed above,
this is particularly sensitive to the kick timescale and geometry and can thus
give unique information on the kick dynamics. While objects with measurements
of all three observables (velocity $v$, initial spin period $P_0$ and alignment
angle $\vartheta_{\Omega\cdot v}$) are rare, the correlations between these are
particularly powerful at testing the models.

Only recently have improved pulsar observations made such comparison possible.
For example, VLBI astrometry provides unprecedentedly high precision proper 
motion measurement \citep[e.g.][]{bri02,cha05}. Also, pulsar timing observations
\citep{hob04,zou05} have recently provided additional (albeit often more uncertain)
proper motion values. Altogether these provide several dozens of useful proper motion
measurements. For pulsar spins, the direction of the projected spin on the sky plane
has been inferred from the radio polarization profile by adopting the rotating vector
model, although uncertainty in the emission mode leads to an inherent $\pi/2$ ambiguity.
Recently, \citet{joh05} obtain the spin orientations of 25 pulsars from radio
polarization observations and noted significant alignment with the pulsar velocities. 
\citet{wan06a} reanalyze archival polarization data and arrive at similar conclusions.
Next, the discovery of X-ray jets and torus structure in PWNe provides an even more
powerful (and largely model-independent) way to obtain the pulsar spin orientation for
a modest number of objects. \citet{ng04} developed a method of pulsar wind torus fitting
to measure the 3-D (not projected!) orientation of the spin axis. So for the vector (2-D)
speeds and spin orientation, progress has been good. In contrast, pulsar initial spin
periods remain very difficult to estimate. Currently, only a handful of pulsars
have initial spin period estimates \citep[e.g. Table~7 in][]{fau06}. As it happens,
most of these have PWN structure indicating their spin orientations, but many do not
have a precise proper motion measurement. In some cases, we can make an estimate of
the space velocity from the structure of the associated SNR, but we must be cautious
in applying such model-dependent values.

To select a sample of pulsars with 2-D velocities, we start from the list of pulsar proper 
motions from \citet{hob05}, selecting young isolated pulsars with $>2$$\sigma$ proper
motion measurements. A few objects are added or updated with new measurements
\citep[e.g.][]{cha05,zou05,ng06}. Galactic acceleration can affect the trajectories
of old pulsars, so we restrict our sample to $\tau < 5\;$Myr (we also have included
the interesting pulsar \object{PSR~B1133+16}, which at $\tau=5.04\;$Myr is just above
this limit). Also, differential Galactic rotation can add appreciably to the observed
space velocity of distant pulsars. When the precise distance is unknown, this can
cause large errors in reducing the speed and position angle to the local standard of
rest. Accordingly, we restrict our sample to $d<3\;$kpc for pulsars with velocities
from astrometric or timing analysis. Table~\ref{t1} lists the pulsar samples used in
this study. The first group of 20 objects have only transverse velocity measurements,
hence they are called `1-D pulsars'. The next group, referred to as `2-D pulsars',
have both a velocity and a projected spin alignment angle. For \object{IC~443} and
\object{PSR~B1800$-$21} the spin orientations are obtained from X-ray PWNe, all other
14 spin measurements are from radio polarization observations \citep{joh05,wan06a}.
As noted, the proper motions, must be corrected to the birth local rest frame for
comparison with the spin directions, as this may affect the position angle by several
degrees.  Finally, the last group consists of 9 `3-D pulsars' which have estimates in
all 3 observables, namely the velocities, initial spin periods and alignment angles,
thus allowing for a full vector comparison. These pulsars place the strongest
constraints on the model parameters. 

We have checked the consistency of the three subsets of pulsars, comparing their
velocity and alignment angle distributions with the Kolmogorov-Smirnov (K-S) test.
The K-S statistics indicates a probability of 0.34 that the velocities of the `1-D'
and `2-D pulsars' are drawn from the same distribution. Comparison of the `3-D'
samples with the summed `1-D' and `2-D' ones gives a similar value of 0.33. For the
alignment angle, the probability of having the same distribution for the `2-D'
and `3-D' pulsars is 0.48. Since the samples are relatively small, these tests are not
very stringent but at least there is no evidence that the different measurement methods 
introduce significant biases in to the sample properties.

We next summarize the status of the several of the particularly interesting `3-D
pulsars'.

\emph{Crab Pulsar (\object{PSR B0531+21})} \citet{ng06} have measured the pulsar's
proper motion with HST astrometry and compared with the spin vector, which was
obtained from pulsar wind torus fitting \citep{ng04}. The initial spin period is
derived from the historical age, measured braking index and spin parameters
\citep{man77}.

\emph{Vela Pulsar (\object{PSR B0833$-$45})} Tracing back the motions of dense
supernova ejecta knots, \citet{asc95} inferred a supernova explosion site
$14\farcm9\pm7\farcm2$ away from the pulsar's current position. With the updated
pulsar proper motion from \citet{dod03}, the kinematic age of the system is then
$\tau=20\pm10\;$kyr. Using the measured braking index of $n=1.4\pm0.2$ \citep{lyn96},
an initial spin period of $30\pm20\;$ms is estimated. PA of the pulsar spin axis is
obtained by torus fitting \citep{ng04}.

\emph{\object{PSR B0540$-$69}} Based on the estimated pulsar age from the SNR expansion
velocity \citep{rey85}, \citet{man93} suggest an initial spin period of 38.7\,\,ms.
The velocity of the pulsar has an initial estimate by \citet{ser04} using HST/WFPC2
images. For the pulsar spin, Ng \& Romani (2007, in preparation) fit the X-ray pulsar
wind torus to obtain its orientation.

\emph{\object{PSR J0538+2817}} \citet{ng07} report a VLBA astrometric measurement
of the pulsar proper motion. They also discovered symmetric structure around the
source in the X-ray observation, which, if interpreted as polar jets or pulsar wind
torus, indicates the spin axis. From the offset of the pulsar position with respect
to the SNR geometrical center, \citet{rom03} argued for a long initial spin period,
which was confirmed by \citet{kra03} with the timing proper motion results and 
strengthened by the precise VLBA proper motion measurement \citep{ng07}.

\emph{\object{PSR B1951+32}} The pulsar proper motion is determined by VLBA astrometry
\citep{mig02}. Assuming the pulsar was born at the center of the shell \object{CTB~80},
the authors deduced the pulsar age and its initial spin period. \citet{hes00} argue that
the H$\alpha$ lobes observed in the HST image correspond to pulsar polar jets, which
indicates its spin orientation. Radio polarization measurements could provide a valuable
test of this hypothesis.

\emph{\object{PSR J0537$-$6910}} The X-ray PWN observation shows bar-like extended
structure around the pulsar. The symmetric structure, if interpreted as pulsar
equatorial outflow, indicates the pulsar spin vector. Ng \& Romani (2007, in preparation)
apply pulsar wind torus fitting to measure the PA of the symmetry axis. The pulsar birth
site may be inferred from the radio image \citep[Fig.~3 in][]{wan01}, where the peak of
the PWN radio emission is at
$\alpha=\mathrm{05^h37^m44\fs5(5) \;,} \; \delta=-69\arcdeg 10\arcmin 11\arcsec(2)$
(J2000). Then, adopting the kinematic age of 5\,\,kyr \citep{wan98}, the pulsar velocity
is $630\;\mathrm{km\;s^{-1}}$. While the initial spin period is unknown, with its
current $P=16\;$ms we can safely assume the pulsar had a $P_0$ range of $8-16\,$ms.

\emph{\object{PSR J1833$-$1034}} \citet{cam06} suggest a young kinematic age
$\lesssim 1000\;$yr for the pulsar, which implies an initial spin period of
$P_0 \gtrsim 55\;$ms. They also estimate the PA of the pulsar spin axis from the
ellipticity of the PWN. Ng \& Romani (2007, in preparation) fitted a model torus
to the X-ray data and obtained similar results. Following \citet{cam06}, we found 
the geometrical center of \object{G21.5$-$0.9} at
$\alpha=\mathrm{18^h33^m33\fs4(3) \;,} \; \delta=-10\arcdeg 34\arcmin 12\arcsec(3)$
(J2000). If this is the explosion site, the offset from the pulsar position gives
an estimated velocity of $125\;\mathrm{km\;s^{-1}}$.

\emph{\object{PSR J0205+6449}} The {\it CXO} images of this pulsar in \object{3C58}
have been noted as having a prominent central torus and polar jets, and fitting of
these provides a robust spin position angle (Ng \& Romani 2007, in preparation).
The position of the pulsar birth may be inferred from the SNR radio expansion center
to be at 
$\alpha=\mathrm{02^h05^m31\fs32(3) \;,} \; \delta=+64\arcdeg 49\arcmin 58\farcs7(5)$
(J2000) (M. F. Bietenholz, private communication). Recently \citet{got07} have
identified a thermal X-ray shell in 3C58 with a similar center. If one assumes that
3C58 is the remnant of SN~1181, one gets an initial pulsar spin of 60\,\,ms
\citep{mur02} and a rather large velocity of $\sim840\;\mathrm{km\;s^{-1}}$.
The X-ray and radio evidence, however, suggest an older age of $3000-5000\;$yr.
This would imply a shorter initial $P_0$ and a transverse velocity of
$\sim 140-230\;\mathrm{km\;s^{-1}}$. While we plot values for the commonly assumed
825\,\,yr age, these values and the pulsar's status as a `3-D object' are certainly
open to question.  A pulsar proper motion would resolve these difficulties.

\emph{\object{PSR B1706$-$44}} A clear equatorial torus and polar jets were found in
{\it CXO} data by \citet{roe05}, where a robust spin position angle and inclination
were fit. This pulsar is often associated with the SNR shell \object{G343.1$-$2.3};
comparing the total energy of the SNR and PWN, \citet{roe05} suggested
$P_0 = 60-80\;$ms. Also, if born at the geometric center of the SNR shell, the pulsar
velocity is $\sim650\;\mathrm{km\;s^{-1}}$. However, the PWN morphology and
interstellar scintillation velocity limits make such a high speed questionable.
Uneven SNR expansion could, for example, make the birth site much closer. Again,
an interferometric pulsar proper motion would be particularly valuable.

\emph{\object{PSR B1800$-$21}} Radio images spanning $\sim 10$ years give an astrometric
proper motion measurement of the pulsar with an accuracy of 2.5 mas~yr$^{-1}$
\citep{bri06}. The Chandra X-ray observation shows symmetric PWN structure around the
pulsar, which is fitted by Ng \& Romani (2007, in preparation) to measure the spin
orientation. However, no initial spin estimate is available for this pulsar.

\emph{\object{PSR J1124$-$5916}} While the PWN image in the initial {\it CXO} HRC
observation showed noticeable ellipticity \citep{hug03}, this axis was not clear
in a subsequent ACIS exposure. A scheduled deep ACIS observation could help to
resolve this structure. A PWN axis would be quite useful here as a robust proper
motion axis and a reasonable age and $P_0$ estimate are available from the highly
symmetric parent SNR.

\section{SIMULATIONS}\label{s4} 

With our pulsar samples defined, we are now ready to constrain the kick models. 
We proceed by making Monte Carlo realizations of a given kick model, drawing
pulsar spin and kick parameters and kick angles from the model distribution,
simulating many pulsar realizations (by integrating according to the equations
in \S\ref{s22}) and projecting onto the sky plane to produce the observables for comparison
with the pulsar sample. The comparison for a given set of true pulsars is made
using a maximum likelihood estimate of the probability of getting that observed
set from the Monte Carlo model.

	Of course, any physical model will have a distribution of values
for the kick parameters rather than single values. What we attempt to extract
from our comparison with the observed pulsars are the characteristic values
in these distributions. For the kick geometry, we assume no preferential directions
in the kick position $\alpha$ and the azimuth angle $\phi$, hence $\cos\alpha$ and
$\phi$ are distributed uniformly between -1 to 1 and 0 to $2\pi$
respectively. The off-axis angle $\theta$ is, however, a more interesting
parameter as it is potentially related to the kick mechanism. Assuming
the kick follows a Gaussian distribution around the surface normal, 
$\theta$ is characterized by the dispersion $\sigma$ in the Gaussian.
Many authors \citep[e.g.][]{hob05} fit the pulsar velocity distribution
with a Maxwellian. We adopt this form for the momentum asymmetry $\eta$,
with characteristic value $\eta_0$. We also attempted to use a simple
Gaussian but this did not give a good distribution for the observed velocities;
apparently, the larger Maxwellian tail is important for matching the observed sample.
For the pre-kick angular velocity $\Omega_i$, we assume a Gaussian distribution with
mean $\bar{\Omega}$. To minimize the fit parameters, we take the dispersion as
$\bar{\Omega}/3$. The value used $\Omega_i$ represents the angular frequency
that the star with the same momentum will have after the proto-neutron star
shrinks to its final radius.
Our toy model approximates the time profile of the momentum kick as being proportional
to the rapidly falling neutrino cooling light curve. A convenient way of introducing a
characteristic kick time scale into the problem is by a delay to the `turn-on' of the
asymmetry, here parameterized as an exponential with a time constant $\tau$. This has
the added advantage of approximating some reasonable physical scenarios for asymmetry
growth. However, as the asymmetry occupies a smaller fraction of the momentum flux in
the light curve tail, the effective $\eta$ must grow. To avoid focus on this strong
correlation, we thus report the equivalent asymmetry $\bar\eta$ here, which
corresponds to the same total thrust for the case of $\tau=0$, i.e.
\[ \bar\eta\int L_\nu (t)\;dt = \eta_0\int \left[1-\exp(-t/\tau)\right]\; L_\nu (t) \;dt \ . \]

In brief, the kick model is characterized by the 4 `fitting parameters':
$\eta_0$, $\bar{\Omega}$, $\tau$ and $\sigma$.
The thrust is given by
\[ F_\nu (t) = \left[1-\exp(-t/\tau)\right]\; \eta \;L_\nu (t)/c \]
and the kick parameters are distributed as
\begin{eqnarray*}
\alpha, \phi & \sim & \mathrm{isotropic} \\
\case{dN}{d\eta} & \propto & \eta^2 \exp(-\eta^2/2\eta_0^2) \\
\case{dN}{d\Omega_i} & \propto & \exp \left [ -\case{(\Omega_i-\bar{\Omega})^2}{2(\bar{\Omega}/3)^2} \right ] \\
\case{dN}{d\theta} & \propto & \sin \theta \exp ( -\theta^2/2\sigma^2) \ . 
\end{eqnarray*}
The factor $\sin \theta$ above accounts for the area in polar coordinates. In the
simulation, the approximation $\sin\theta \approx \theta$ is taken to simplify the
calculations. Also we tabulate the asymmetry results as $\bar\eta$, as noted above.

Given a set of fitting parameters, the kick velocity and spin vectors of
$N_{\mathrm{sim}}=2\times 10^5$ pulsars are simulated. As the pulsar radial velocity
is not an observable, the data samples are essentially only projected 2-D values on
the sky plane. Therefore, the simulation results have to be projected before comparing
to the observations. This is done by specifying 2 angles: the inclination angle $\zeta$
of the spin axis to the sky plane, and the azimuth angle $\phi_v$ of the velocity about
the spin vector. While the latter is totally random, $\zeta$ is sometimes actually known
from pulsar wind torus fitting. In the simulation, each of the simulated vectors are
projected $N_\mathrm{proj}=50$ times, according to the measured $\zeta$ (and its error)
of an individual pulsar in the sample if this is known, or isotropically otherwise.

In order to compare the model prediction to measurements, we define a likelihood
function similar to the typical $\chi^2$ statistic
\[ \mathcal{L}^{ij}_\xi \sim \exp \left [
-( \xi^i_\mathrm{obs} -\xi^j_\mathrm{sim})^2 /2(\sigma_\xi^i)^2  \right ] \ .\]
For an observable $\xi \in \{v,P_0,\vartheta_{\Omega \cdot v}\} $,
$\xi^i_\mathrm{obs}$ is the $i$-th sample pulsar's measurement with uncertainty
$\sigma_\xi^i$ and $\xi^j_\mathrm{sim}$ is the $j$-th simulation result. For the
initial spin periods $P_0$ and velocities based on the DM-estimated distance,
uncertainties in the measurements are dominated by systematic errors which do
not follow Gaussian properties. In this case, a boxcar function is used
\[ \mathcal{L}^{ij}_\xi = \left\{ \begin{array}{ll}
1 & \mbox{if \, $\left |\xi^i_\mathrm{obs}-\xi^j_\mathrm{sim}\right |\leq\sigma^i_\xi$} \\
0 & \mbox{otherwise}
\end{array} \right. \ . \]
For each pulsar in the data sample, the likelihood function is then the product of all
its observables, averaged over the simulations and projections 
\[ \mathcal{L}^i = \frac{1}{N}\sum_j^{N} \left ( \prod_{\xi} \mathcal{L}^{ij}_\xi \right ) \ , \]
where $N=N_\mathrm{sim} \times N_\mathrm{proj}$. Finally, the total likelihood function
is the product of all the pulsars in the data sample
\[\mathcal{L}= \prod_i^{N_\mathrm{obs}} \mathcal{L}^i \ . \]
It is convenient to define a Figure-of-Merit (FoM) function for comparison between
different models. As usual, it is defined as the negative log of the likelihood
function FoM$=-\log \mathcal{L}$. 

\section{Results\label{s5}} 

\begin{figure}
\plotone{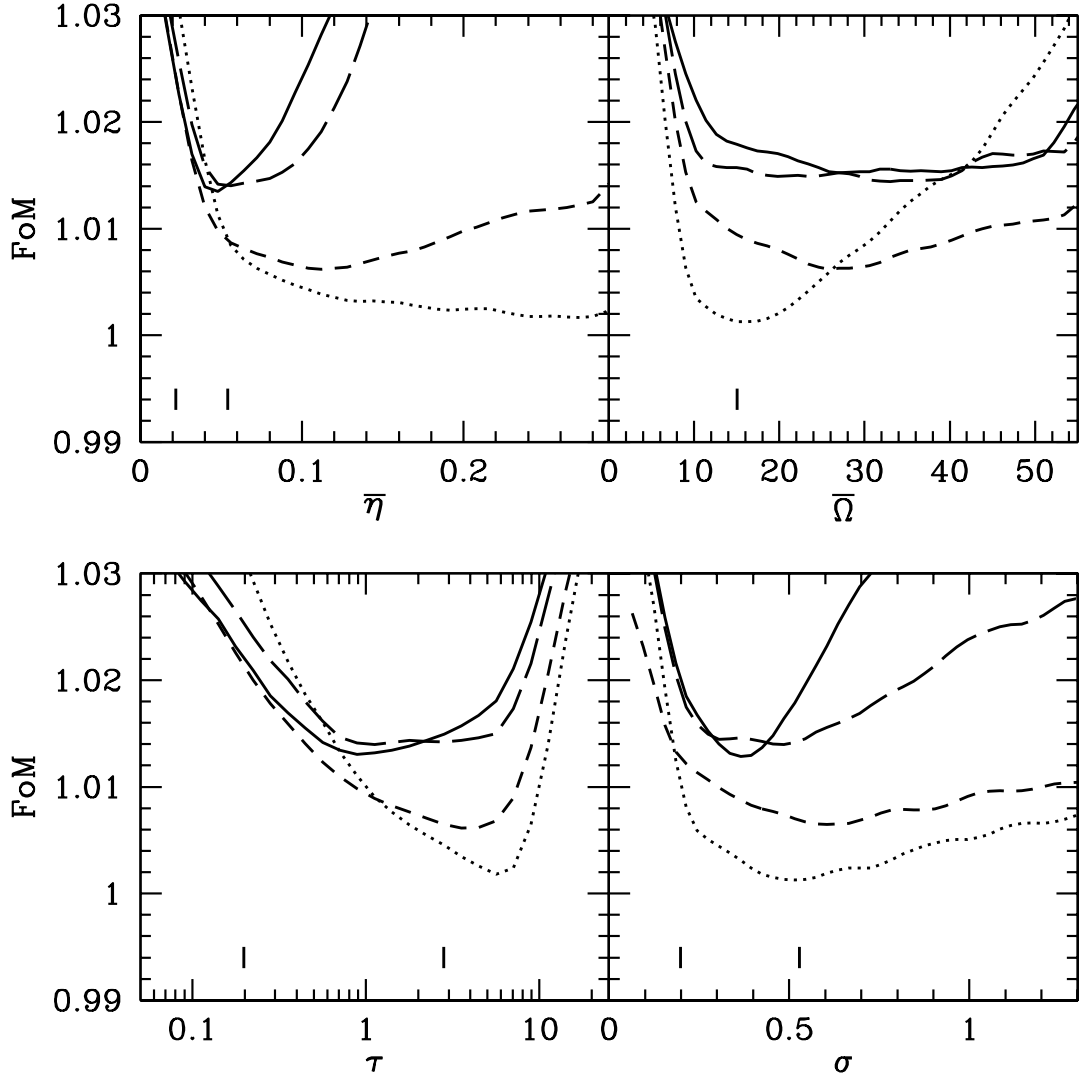}
\caption{Projected FoM for individual parameters of the `quasi-stationary' model.
The solid lines show the full sample fit; the long dashed lines show the fit with the
`1-D, 2-D pulsars' and the `minimal set' of `3-D objects' (i.e. Crab, Vela and B0540$-$69);
the short dashed lines show the fit with only `1-D' and `2-D' pulsars; the dotted lines
show the fit with only `1-D pulsars'. Smoothing filters have been applied to the curves
and the FoM scale has been normalized by the minimum `$\nu$-transport' value for each
data set. For the full sample curves, the upper and lower `1$\sigma$' intervals provided
by the bootstrap analysis (Table~\ref{t2}) are indicated by the tickmarks.
Comparison with the $\Delta$FoM at these confidence intervals and with Table~\ref{t2}
allows the reader to judge acceptable parameter ranges.
\label{f5}}
\end{figure}

\begin{figure}
\plotone{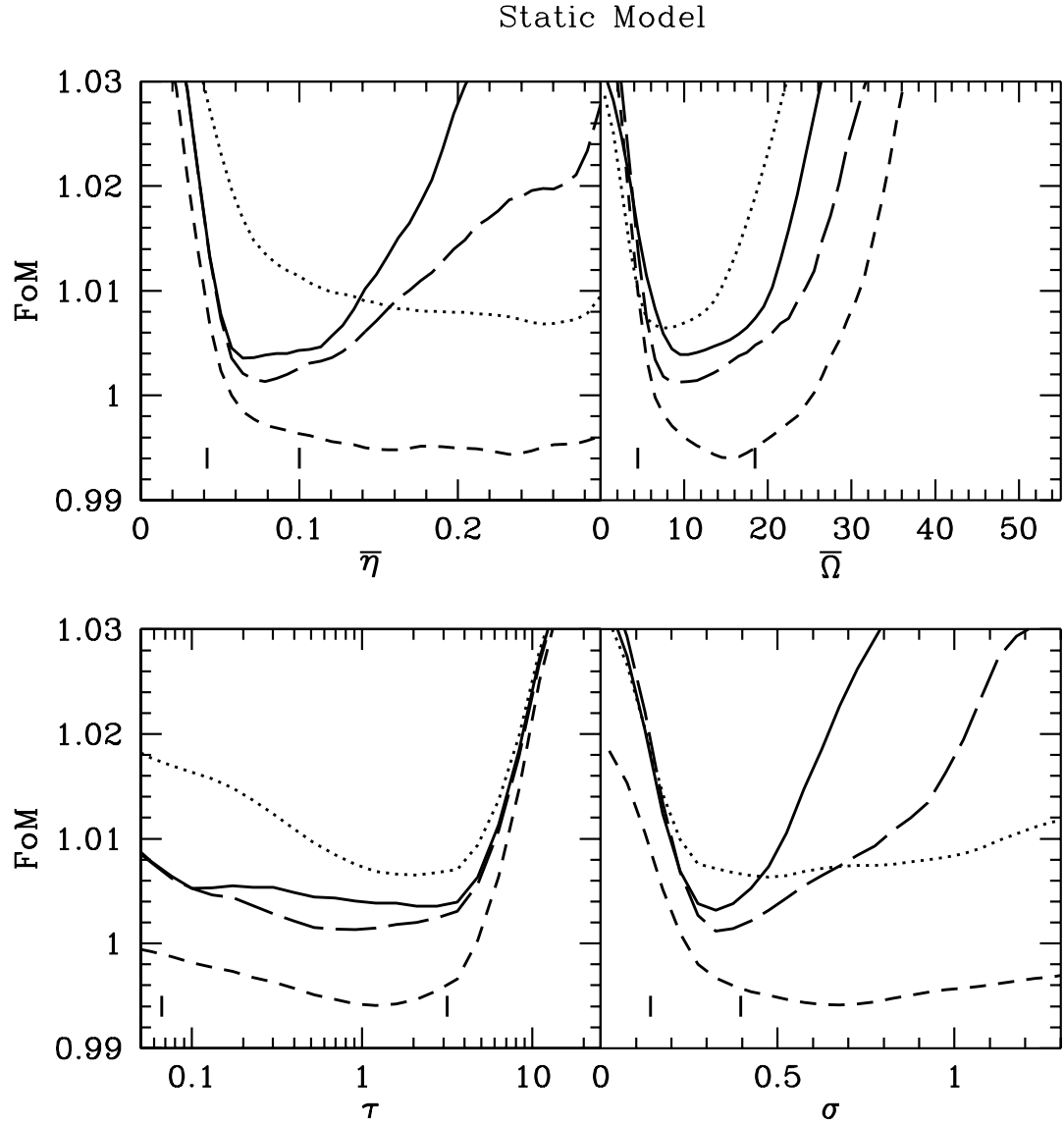}
\caption{Same as Fig.~\ref{f5}, for the `static' model.
\label{f6}}
\end{figure}

\begin{figure}
\plotone{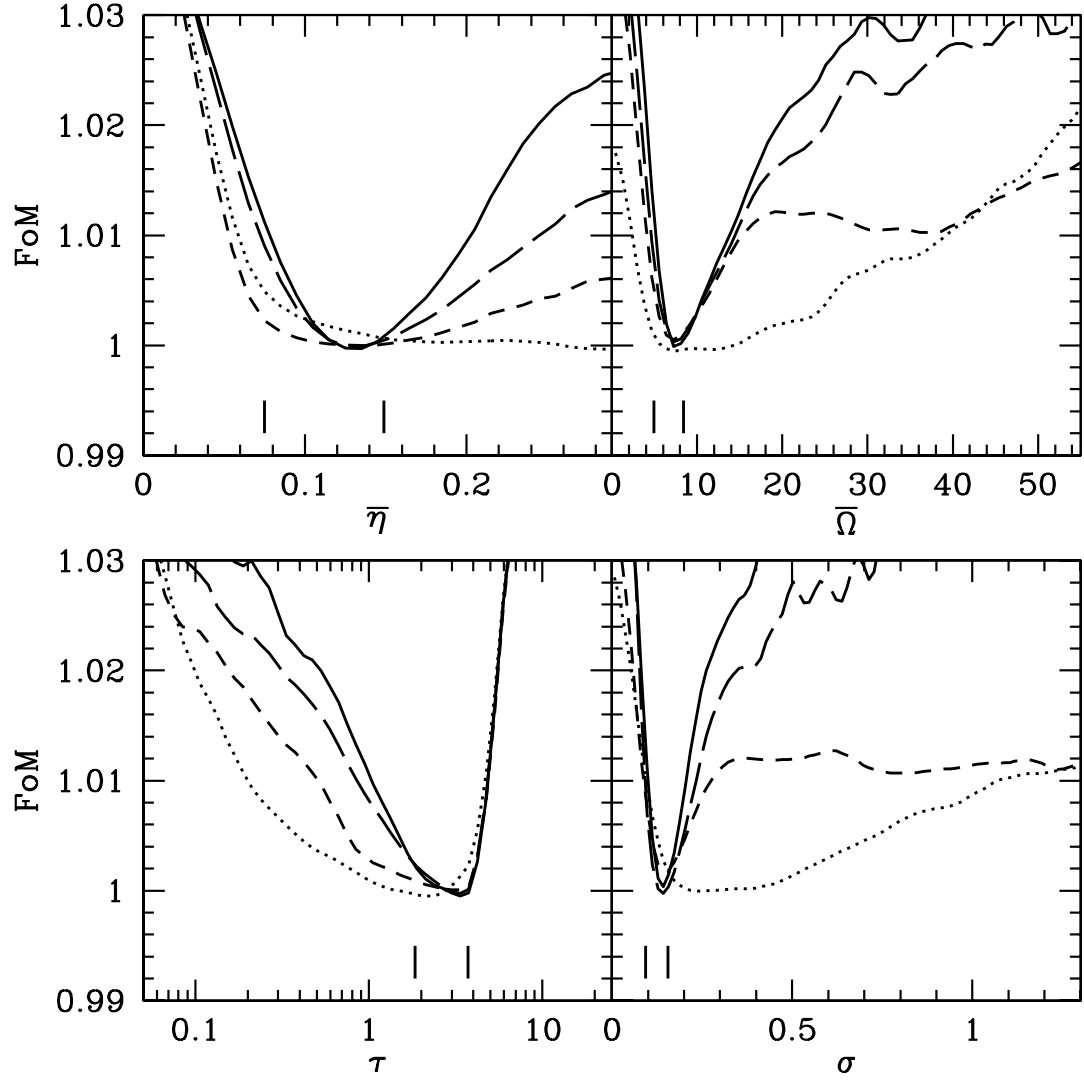}
\caption{Same as  Fig.~\ref{f5}, for the $\nu$-transport model.
\label{f7}}
\end{figure}

Figures~\ref{f5}$-$\ref{f7} show the projections of the likelihood function of our
three kick scenarios for the main model parameters.  Each panel shows four curves,
with the minimum indicating the best-fit parameter values. The solid curve shows
results obtained using all the objects in our pulsar sample; this generally has the tightest,
best-defined minimum. However, as discussed in \S\ref{s3}, some of the spin and velocity
estimates for certain `3-D objects' are quite model-dependent and can be questioned.
We therefore also ran the FoM calculations with a `minimal' set of `3-D pulsars':
the widely accepted Crab and Vela pulsars and PSR~B0650$-$69, which is simple, but
has large statistical error bars. The results are very similar to those for the 
full set, albeit with larger uncertainties. Finally, we ran fits with the `1-D' and 
`2-D' objects together and just the `1-D objects' alone.  Again, the fits are broadly
consistent with the full set's results, but as expected the minima are poorly
constrained. Thus we confirm that our handful of well measured `3-D objects'
dramatically improve our knowledge of the birth kick properties, but apparently do
not introduce major biases into the fits.

	Our likelihood function is unnormalized. Since it is a product over individual
objects, it is convenient to rescale the values for the curves computed for different
sets of pulsars. This removes the gross scaling of FoM with number of objects and
facilitates cross-comparison.  In practice the results with different object sets are
normalized by dividing all by the corresponding best-fit FoM values for the
$\nu$-transport model for the same data set. Note that, with the exception of the
`1D+2D' (dashed line) fit to the static model, all minima occur at FoM $>1$. 
Also the `$\nu$-transport' model has narrower minima. This
means that this model, in addition to being the most physical,
provides the best description of the full data set of the three scenarios tested.

It is useful to understand how the `1-D objects' can provide some model constraints,
i.e. how the pulsar velocity distribution is sensitive not only to the asymmetry
$\eta$, but to the other kick parameters, as well. To illustrate, consider the case
of large $\sigma$ or small $\bar\Omega$, i.e. when kick-induced spin is important.
When the pulsar is spun up, the induced linear and angular momenta are always
orthogonal. In this case, the rotational averaging is efficient, generating many low
velocity pulsars.  In contrast, for large $\bar\Omega$ or nearly normal kicks
(i.e. small $\sigma$), the initial spin sets the averaging axis, and the final velocity
is principally determined by the kick location. For example, a polar kick results in a
fast moving pulsar, while an equatorial kick is quickly averaged to give a low final
speed. Thus, for fast initial spinners, the kick velocity distribution becomes
relatively uniform, as our model assumes no preferential kick direction. This is
unlike the observed velocity distribution.

\begin{figure}
\includegraphics[scale=0.75]{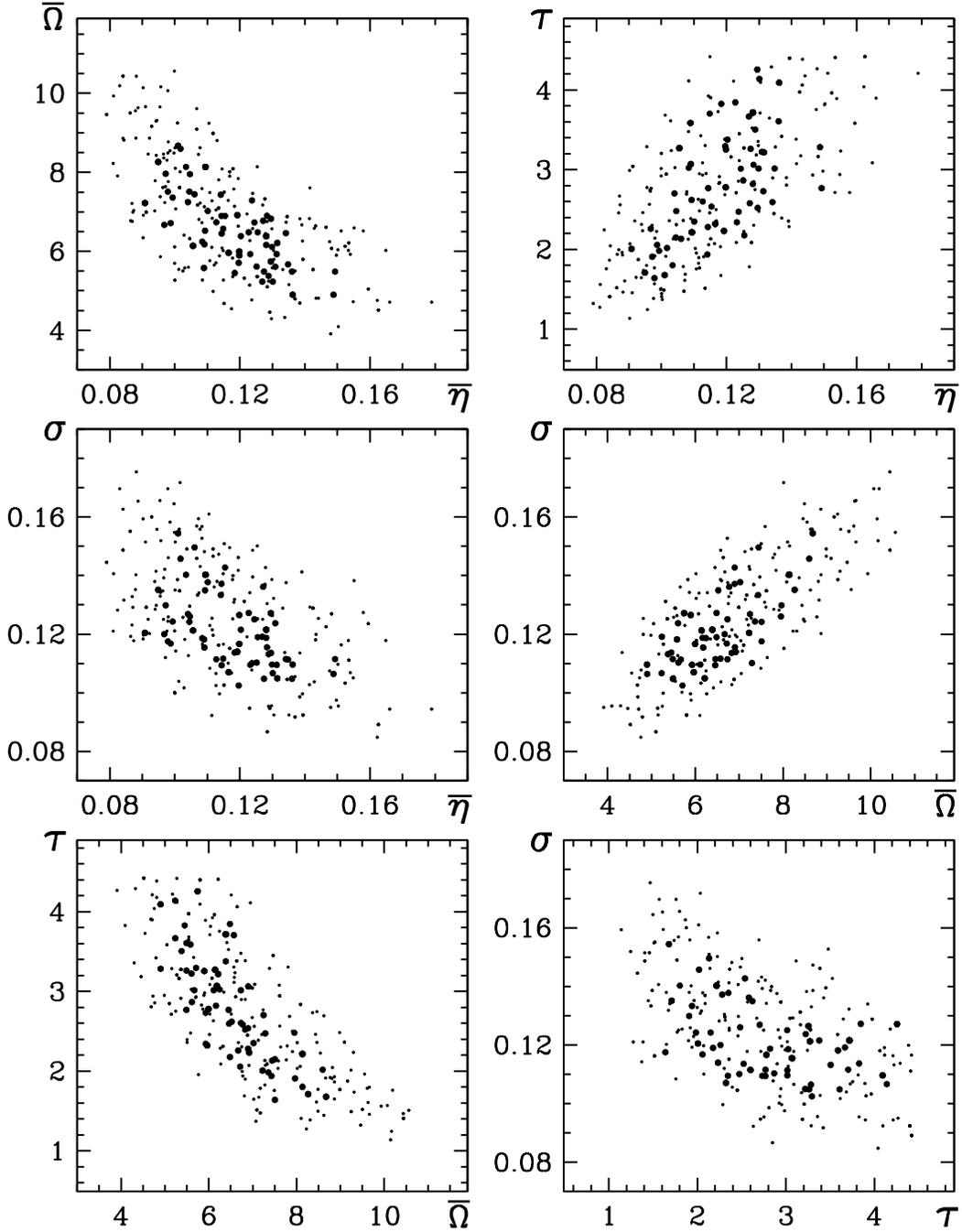}
\caption{Correlations between fitting parameters for the best-fit $\nu$-transport
model with full data sample. Large and small dots are simulations within 1 and
$2\sigma$ of the best fit FoM. Note that significant correlations remain between
the parameters and that our full projected errors are (conservatively) larger than
the single parameters uncertainties.
\label{corr}}
\end{figure}

	Of course, when only velocity values are available, there is a large co-variance between
the model parameters, leading to a broad minimum in the multi-dimensional (projected)
FoM curve. In fact, it is the handful of `3-D objects' that break these degeneracies and
best localize the minima, as the FoM curves illustrate. Even then, there are significant
correlations in the fit parameters, as illustrated by Fig.~\ref{corr}. The points show
the loci of the minimum in two dimensional cuts through parameter space for the
$\nu$-transport model.  Similar results are obtained for the other models.
To be conservative, we quote here the `projected' multi-dimensional errors
on all the model parameters.  The single-parameter errors are, in general,
significantly smaller.
Table~\ref{t2} reports the best-fit parameters for the models, with $1\sigma$ 
confidence intervals estimated from the bootstrap percentiles \citep{efr93}. 
We generated $10^5$ bootstrap samples and quote $1\sigma$ errors from the
standard normal 68\% interval of the bootstrap replica distribution. The two dot sizes
in Fig.~\ref{corr} show the loci of the `$1\sigma$' and `$2\sigma$' minima.

	There are some robust trends in the fitting results. All four of our basic model
parameters are significantly constrained and, with the exception of $\tau$ in the
`static' model, they have significantly non-zero values.
Thus, we conclude that in all of our scenarios a finite pre-kick spin and a finite
tangential component to the kick are required. In some sense, we consider the
$\nu$-transport model to be the most physical. The best parameters for this model are
closer to those of the static model than to the shrinking `quasi-stationary' model.
Thus, the kick is somewhat insensitive to the initial envelope shrinkage of the
$\nu$-transport model. However, it should be noted that the $\nu$-transport model
strongly prefers a finite $\tau \sim 3$s. This {\it ensures} that the kick is
insensitive to the initial large envelope phase. Interestingly, if $\tau$ is forced
to be small for this model the best fit $\bar\eta$, $\sigma$ and $\bar \Omega$ move 
much closer to the preferred `quasi-stationary' values (Fig.~\ref{corr}). In this limit,
the kick is dominated by the large-radius initial phase of the cooling curve.

The final angular velocity (after the proto-neutron star cooling and kick) has 
a very broad distribution, which peaks at $\sim120\;\mathrm{rad\;s}^{-1}$ for 
the $\nu$-transport model.  Similar results are obtained for the other models.
With the pre-kick spin $\Omega_i$ of only $10-30\;\mathrm{rad\;s}^{-1}$, this
indicates that most ($> 90\%$) of the spin is induced by the kick, as suggested 
by \citet{spr98}.
Unfortunately, so few initial spin estimates are available at present that the
post-kick spin distribution cannot yet be compared with the data.

\begin{figure}
\plotone{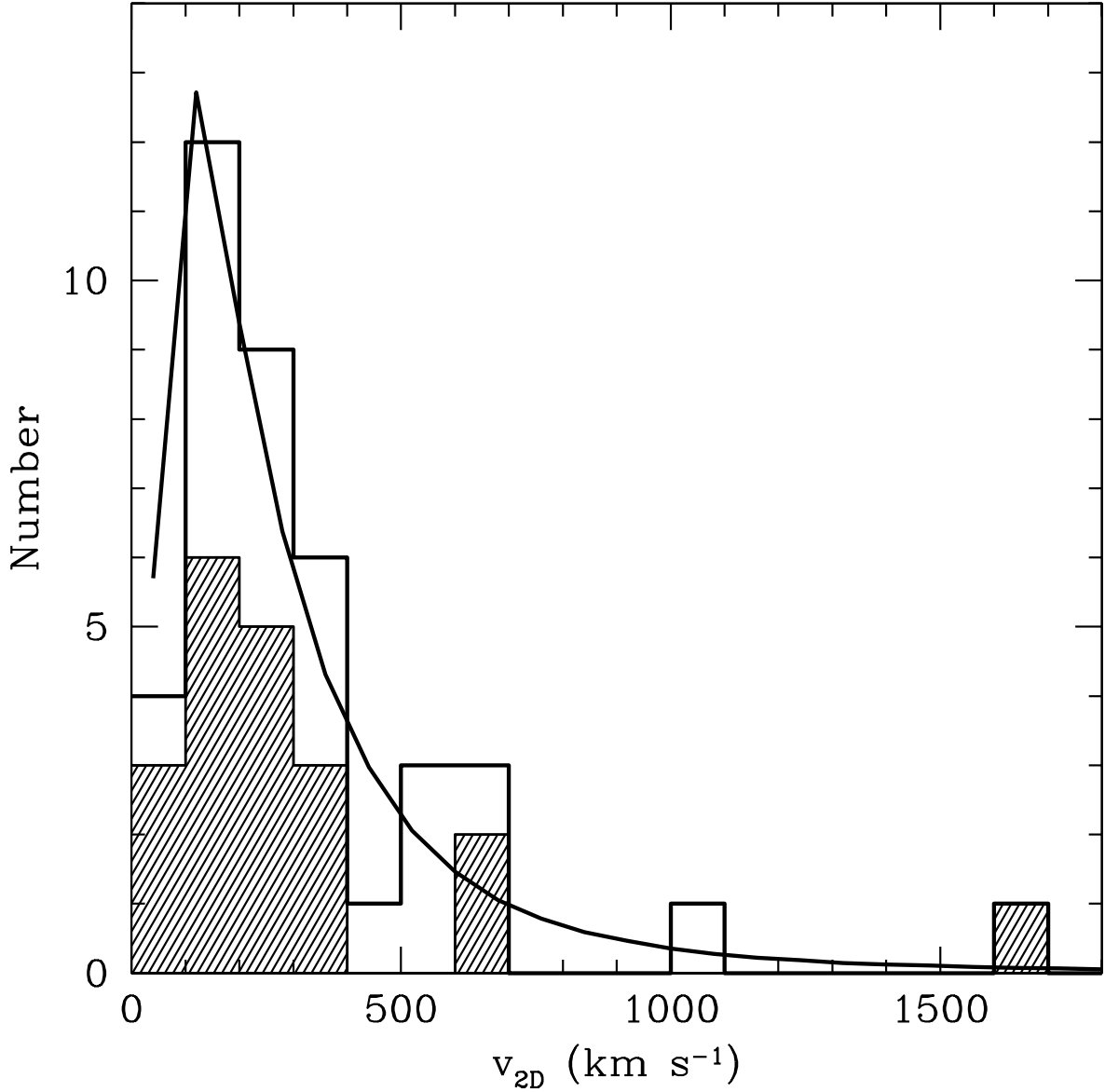}
\caption{Velocity distributions of the best-fit $\nu$-transport model with the full
data set, compared to the observed pulsars with proper motion measurements of
$>2\sigma$ significance. The `1-D pulsars' are shaded.
\label{histv}}
\end{figure}

\begin{figure}
\includegraphics[angle=270,scale=0.6]{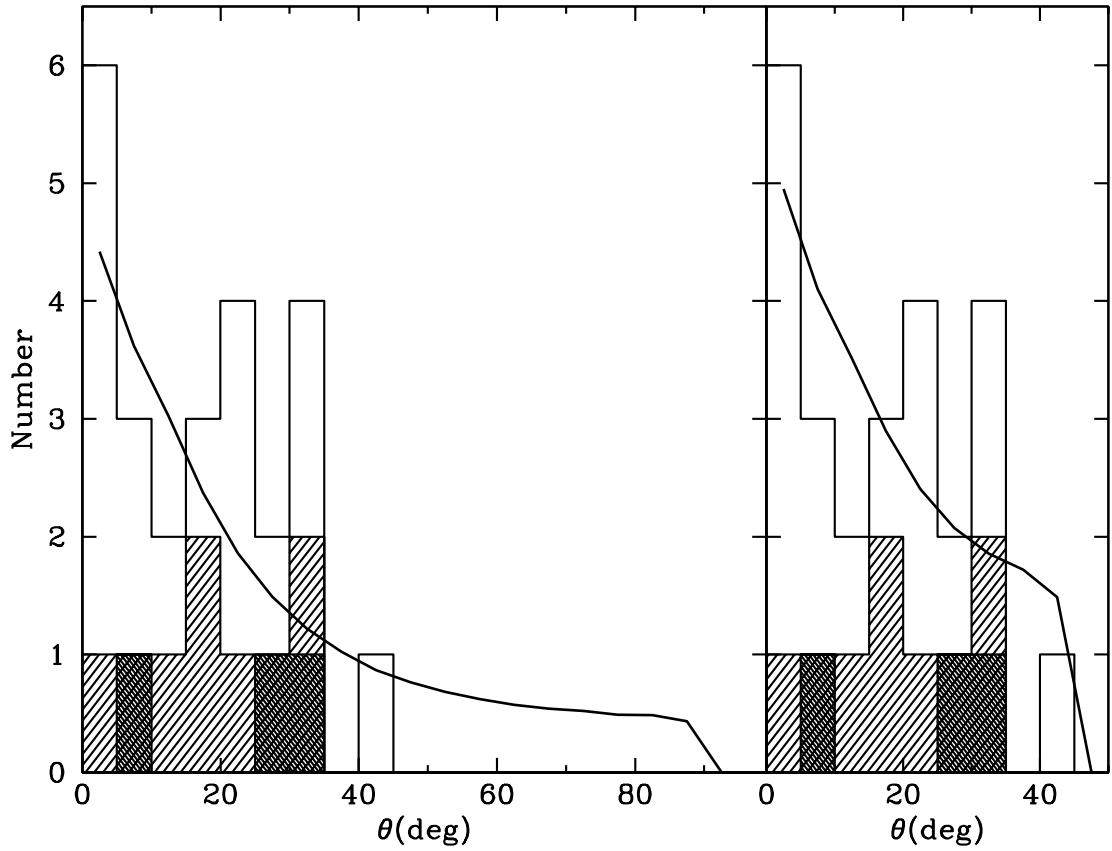}
\caption{Alignment angles of the pulsar samples compared to the best-fit $\nu$-transport 
model with the full data set. The `3-D objects' and the `minimal set' are
shaded in light and dark gray respectively. \emph{Right panel:} Simulations are folded
around 45\arcdeg\ to illustrate the effects of the 90\arcdeg\ ambiguity in
alignment angles measured by radio polarization.
\label{histt}}
\end{figure}

The distribution of final velocities for the best-fit $\nu$-transport model
constrained by the full data sample is plotted in Fig.~\ref{histv}. The observed velocities represent
reasonable draws from that model, with a K-S probability of 0.3, although the model
predicts some additional low velocity pulsars. Similarly,
Fig.~\ref{histt} shows the model and data distributions of the alignment angles.
The left panel compares the model angle distribution and the data. This is the
alignment angle range that is actually used in the fits. The K-S statistic gives a moderately
poor fit with a probability of only 0.08 that the data and model have the same
distributions. Note, however that the `2-D pulsars' with angles determined by
polarization have, by fiat, $\vartheta < 45\arcdeg$. In the right panel we fold
the model curve back to $\vartheta < 45\arcdeg$. The data match this distribution
very well, with the K-S probability increased to 0.98. If our model is correct,
about 20\% of the `2-D' $\vartheta$ values should be $>45\arcdeg$. Note also that a
few of the `3-D' measurements have large errors and values $>45\arcdeg$ are allowed.
We have tested this by randomly drawing $\sim$3 of the `2-D pulsars', flipping the
table values to $\vartheta = 90\arcdeg-\vartheta$, and re-fitting. The majority of the
draws produce, as expected, best-fit parameters close to our adopted values, but deeper
and tighter minima.
%EXAPND ON THIS A BIT>>>>>

\section{Discussion} 

We have explored the effect of an idealized model of a pulsar kick: a single finite
duration thrust applied at a random but fixed point on the
surface of a rotating, cooling proto-neutron star. This is a reasonable direct
description of a class of models where neutrino anisotropy is imposed by a structure
(e.g. magnetic) locked in the underlying neutron star. It is also a reasonable
stand-in for other physical models where e.g. convection is controlled by a hot spot
on the star surface. We note that our modeled momentum thrust can be considered as the
net anisotropy of any number of kicks, as long as the anisotropy maintains a fixed
pattern and follows the evolving core luminosity. Our models do not directly address the
effects of multiple, completely random thrusts, but as first described by \citet{spr98}
and as confirmed in recent model integrations of \citet{wan06b}, multiple kicks
tend to wash out any anisotropy; the net effect of a large number of kicks is
to produce little or no correlation between motion and spin, contrary to recent
observations. Our simulations also do not directly test the gravity-driven kicks described by
\citet{sch06} since for a rotating star these are almost certainly directed
at the pole. Their scenario also does not include the kick-induced spin and rotational
averaging described here. The technique described here for confronting
parameterized models with the pulsar data set can, and should, be applied to 
formulations that probe such kick physics. However, we expect that the basic
parameter estimates here, describing kick direction timescale and magnitude, can
be a useful guide for evaluating a wide range of models.

\begin{figure}
\includegraphics[angle=270,scale=0.7]{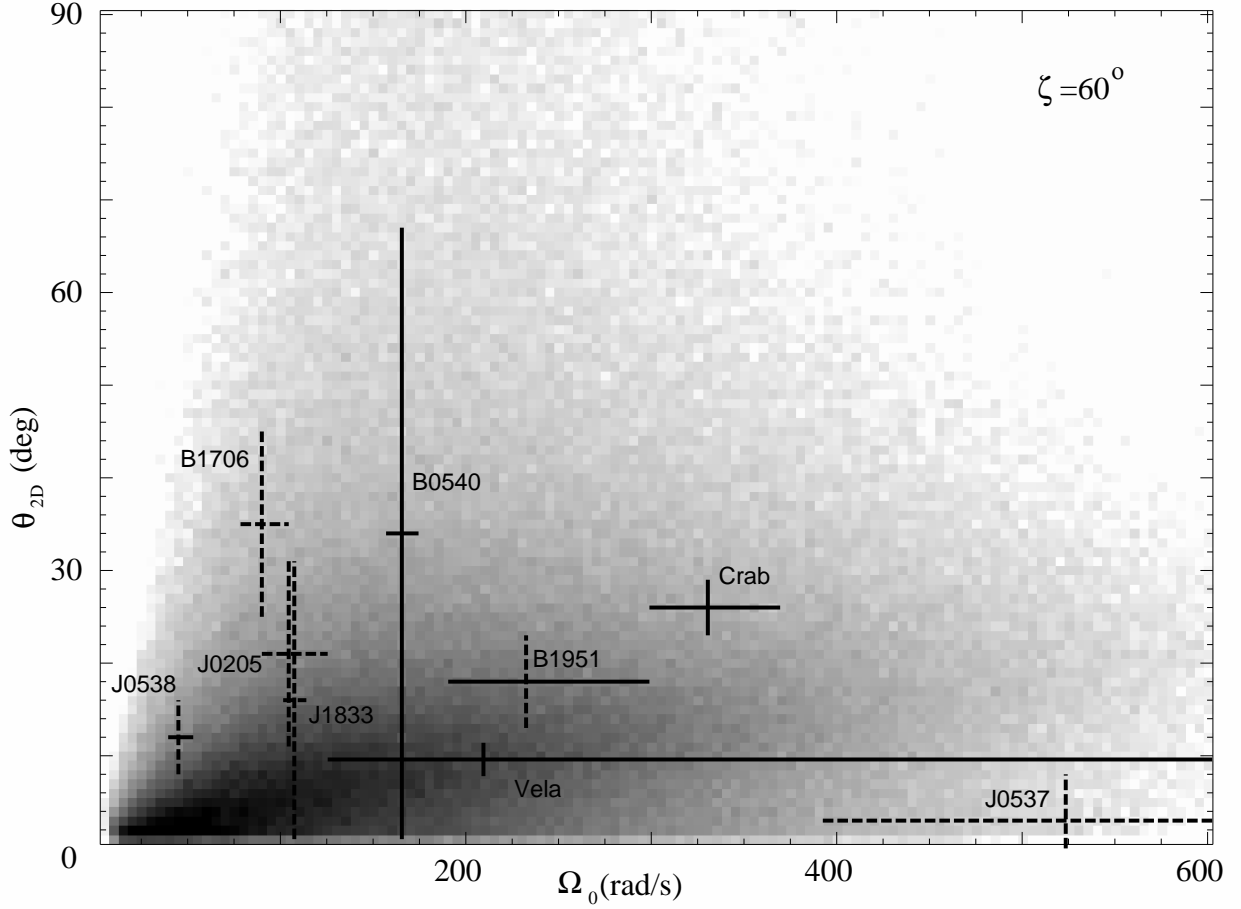}
\caption{Distribution of alignment angle $\vartheta_{\Omega\cdot v}$ vs.
post-kick initial spin $\Omega_0$ for the best-fit
$\nu$-transport model with the full data set. The simulations are projected for
$\zeta=60\arcdeg$, a typical value for the `3-D pulsars'. For these pulsars,
measurements with model-dependent assumptions dominating the errors are shown by
plotting the corresponding error flags in dotted lines.
\label{thw}}
\end{figure}

\begin{figure}
\includegraphics[angle=270,scale=0.8]{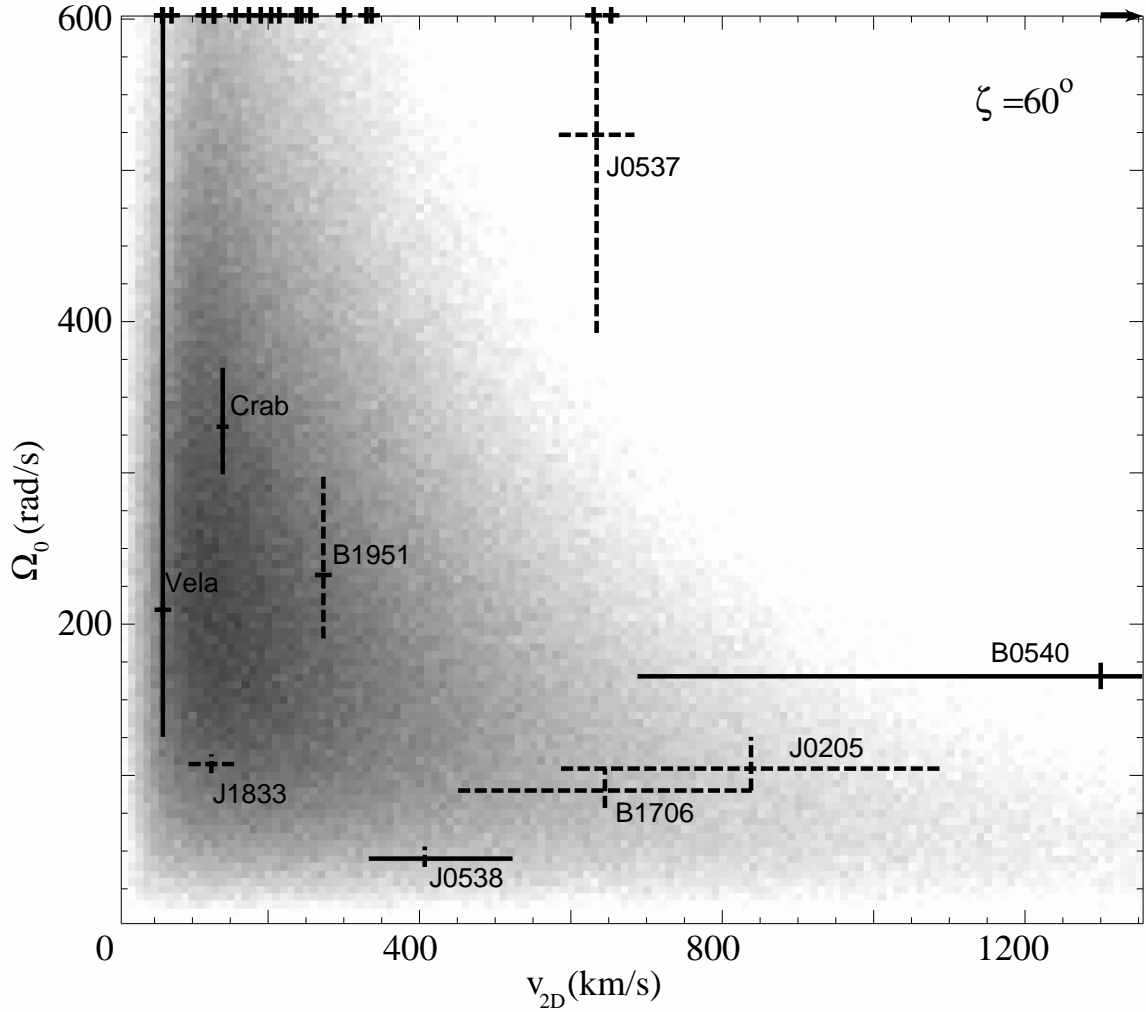}
\caption{Distribution of pulsar post-kick initial spin vs. velocity for the best-fit
$\nu$-transport model with the full data set. The simulations are projected for
$\zeta=60\arcdeg$. The velocities of the `1-D pulsars' are marked at top. 
Again, systematic-dominated parameters of `3-D objects' have dotted-line error bars.
\label{wv}}
\end{figure}

	While our model is certainly a simplified parameterization of real physics, it
does produce some surprisingly complex behavior that can be discerned in the real
pulsar data. The most interesting effects are only visible in the `3-D pulsars',
objects for which we have multiple observables and can measure their correlation
within the population. Consider, for example, the projected angle
$\vartheta_{\Omega\cdot v}$ vs. the post-kick initial pulsar spin frequency
$\Omega_0$ (Fig.~\ref{thw}). The model distribution shows that at intermediate initial
spin periods $40-60\;$ms, we expect a substantial fraction of pulsars to be misaligned,
with $\vartheta_{\Omega\cdot v} > 45\arcdeg$. This is because during the kick, the spin
was slow enough that the initial impulse induced the bulk of the spin, at orthogonal
angles. In contrast, short spin period pulsars ($P_0 < 20\;$ms, $\Omega_0 >
200\;\mathrm{rad\;s^{-1}}$) tend toward alignment due to spin averaging, and very long
period pulsars ($P_0>100$\,ms, $\Omega_0 < 60\;\mathrm{rad\;s^{-1}}$) are also aligned.
The latter effect is `survivorship'; to retain a slow spin even with a decent kick
the pulsar must have started as a slow spinner {\it and} have been kicked nearly
radially. Similar correlations exist with the pulsar
velocity, e.g. the fastest moving pulsars are expected to be slow spinners
(Fig.~\ref{wv}): if the initial kick produces a large spin, then continued thrust
`averages away' and the pulsar never reaches high speeds.

\begin{figure}
\includegraphics[angle=270,scale=0.6]{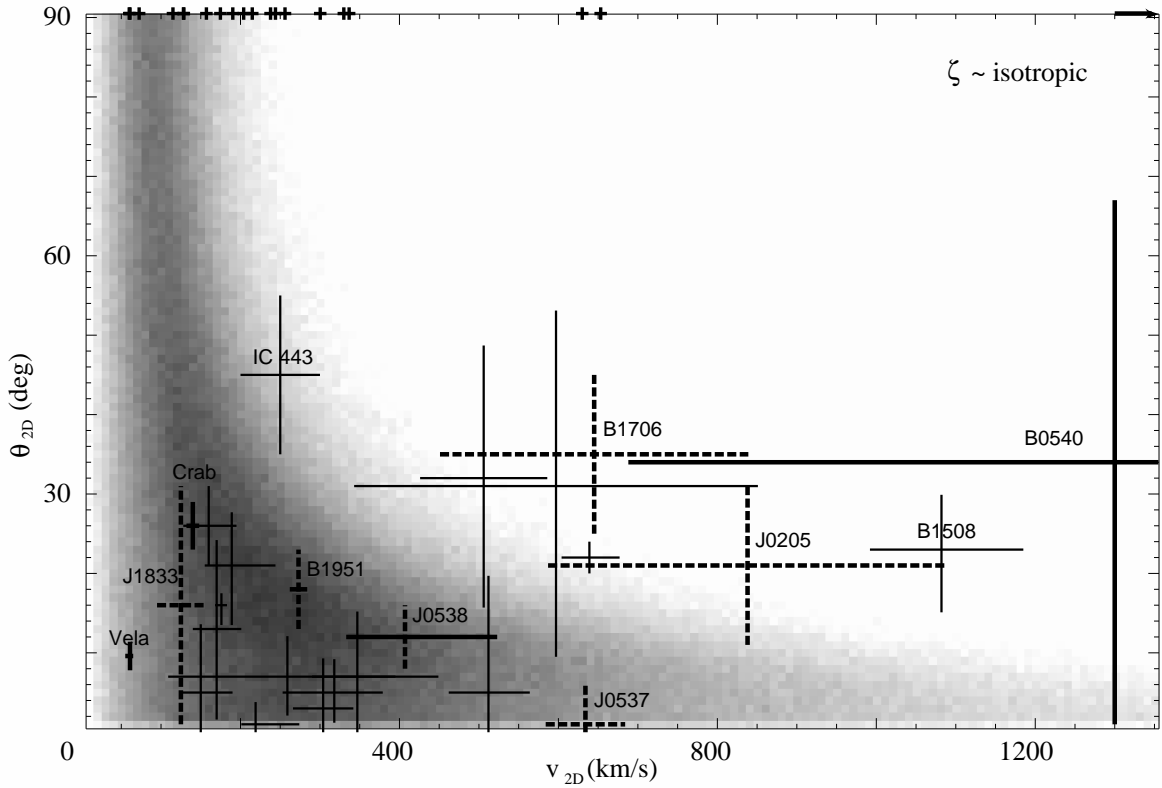}
\caption{Distribution of alignment angle vs. kick velocity for the best-fit
$\nu$-transport model with the full data set. Here we can plot many `2-D pulsars', so
the simulations are projected isotropically. The velocities of the
`1-D pulsars' are plotted on the top. `2-D objects' are plotted by the thin lines.
Measurements of `3-D objects' have thick-lined error bars, with dotted lines
for those parameters dominated by model-dependent systematics.
\label{thv}}
\end{figure}

	While considering the 2-D correlations, it is interesting to look at the
alignment-velocity correlation (Fig.~\ref{thv}). Here we may include many
of the RV-model pulsars. It is not clear that these objects follow the expected
correlations. First, there appears a dearth of misaligned pulsars at slow speeds.
However, two effects may explain this. First, truly slow movers are selected against
in the data sample as they are unlikely to have either a significant astrometric or
timing proper motion.  Second, we should remember that all RV pulsars have been
{\it forced} to have $\vartheta_{\Omega\cdot v} < 45\arcdeg$. It is possible that at least
a few of the slower pulsars are in fact orthogonal. What about the pulsars at large
space velocity in this plot?  Well, the model suggests that these should show good
alignment. Two of the `3-D pulsars' PSRs B1706$-$44 and J0205+6449 in 3C58 show
surprisingly large $\vartheta_{\Omega\cdot v}$ for their speed. However, note that
the speed plotted for PSR~J0205+6449 assumes a birth date of CE~1181. If, as appears
increasingly likely \citep{got07}, the pulsar is actually $\ge 3,000$ years old, then
the object falls near the peak of the expected distribution. Similarly, the large
velocity inferred for PSR~B1706$-$44 depends on travel from the present remnant center.
Scintillation measurements suggest a much slower speed, requiring an asymmetric SNR or
even non-association; these again put the pulsar near the peak of the model distribution.
Further note the large error bar on the alignment angle of PSR~J1833$-$1034 in
G21.5$-$0.9. This is set by the uncertainty in the SNR expansion center
position. It would not be very surprising if this object, as well as the neutron star
\object{CXOU~J061705.3+222127} in IC~443, would appear at $\vartheta_{\Omega\cdot v} > 45\arcdeg$.

Recall that for the 2-D polarization objects, set to $\vartheta_{\Omega \cdot v} <
45\arcdeg$
by fiat, we might expect that 3-4 are actually at large $\vartheta$, emitting in the
opposite polarization mode.  This would not, of course, significantly dilute the
general trend noted by e.g. \citet{joh05}. that RV model fits correlate with proper
motion vectors. It is not clear if we will be able to understand the radio emission
well enough to remove this $\pi/2$ ambiguity. Happily the direct X-ray PWN fits do
not suffer this problem, so more such measurements should expose a few poorly aligned 
objects. Globally, we expect fast pulsars to be well aligned. Slow pulsars will have
a significant fraction of poorly aligned objects. Note that additional velocity
components, e.g. orbital velocity for pulsars released from binaries, will enhance
this trend.

	Are there any serious challenges to our proposed model? Well, yes.
\object{PSR~B1508+55} is fast and has a relatively large $\vartheta_{\Omega\cdot v}
= 23\arcdeg\pm7\arcdeg$. This is very difficult to produce with the sort of kick
scenarios that we describe. We would predict that improved spin measurements would
prefer smaller $\vartheta < 15\arcdeg$. If a large misalignment perseveres, some
amendment is needed to the kick scenario. 

	What about the substantial velocities which may arise naturally due 
to convection, when $m=1$ modes dominate \citep{sch06}? These kicks make
{\it vector} predictions differing from those modeled here. For example, these
authors suggest that fast spinners produce strong $m=1$ modes and thus only these attain
high space velocities (opposite to the pattern of Fig.~\ref{wv}). Also, they suggest
that slow spinners ($ \lesssim 60\;\mathrm{rad\;s^{-1}}$) should be poorly aligned
(cf. Fig.~\ref{thw}). This picture may also produce a bi-modal distribution
of pulsar speeds; it is presently unclear if such bi-modality exists, but this
would be a important test of the model. Perhaps the most interesting {\it vector}
test of this down-flow kick scenario should come from binary pulsars, where a
rotation-controlled $m=1$ mode predicts a kick along the binary orbit normal. 
We have not discussed binary kicks in this paper, deferring to a future publication.

	Also, we should at least comment on the physical plausibility
of our fit values. The effective asymmetry $\bar \eta$ is quite large.
If this were strictly induced at the neutrino photosphere, this would
represent a large asymmetry. In the simple magnetic field picture this
requires very large (although possibly transient) fields of $\sim 10^
{15.5-16}\;$G. Thus, a momentum asymmetry mediated by coupling to matter
would be quite appealing, as the field requirements are lower. The pre-kick
spin periods are, as noted above, generally quite low. This argues for
effective core-envelope coupling, possibly magnetic, in the pre-collapse
progenitor \citep{spr98}. What about the modest $\sigma \approx
0.2-0.4 \approx 10\arcdeg-20\arcdeg$ angle of the kick to the normal? This
does seem quite natural in a magnetically-induced asymmetry model,
where a dipole centered at $\sim R_{NS}\sigma/2$ would cause such an angle.
Pictures where the kick is due to matter fall-back, would tend to
accrete matter with a large specific angular momentum. While
effective at re-spinning the star, this would tend to have a large
$\sigma > 1$ and would, of course, suggest a strong trend to 
$\vartheta_{\Omega \cdot v} \sim 90\arcdeg$, which are not observed. 

	Finally, we come to the time delay in the kick asymmetry. Since
the initial neutrino flux falls off exponentially, the delay $\tau$
can be thought of as selecting a characteristic timescale for the
momentum thrust. As already noted, this has a strong covariance with
the kick amplitude and with the pre-kick spin $\Omega_i$, with small
delays requiring smaller kicks and faster spins. For example the best-fit
`static' model is consistent with no delay and has
$\Omega_i \approx 10-15\;\mathrm{rad\;s^{-1}}$.
The quasi-stationary model prefers a $\sim 1$\,s delay and has 
$\Omega_i \approx 20-40\;\mathrm{rad\;s^{-1}}$.
Interestingly all the fits to proper motion
data alone prefer some substantial delay, implying that long kicks and significant
rotational averaging are useful in producing the observed velocity distribution.
Finally, the $\nu$-transport model very strongly excludes the no-delay solution. As
noted above, this is required to avoid having the bulk of the momentum thrust while
the star still has a large radius. What about a physical origin for this delay? In a
picture where a large $B$ field introduces the asymmetry, it is attractive to associate
this field with early dynamo action in the proto-neutron star. Such dynamo theory is not
well developed, but it is interesting that qualitative discussions \citep[e.g.][]{tho93}
suggest a characteristic time of $\sim 1-3\;$s for a convectively-driven dynamo to drive
the field to saturation. If $\eta$ grows as $\sim 0.1B_{15} (MeV/\bar{E}_\nu)^2$, then
the kick gains in efficiency as the $B$ field grows to its limit and as the
neutrinosphere temperature drops. Thus, the preferred fit parameters, a delayed kick with modest
tangential component and a slow-spinning pre-kick star, are quite consistent with our
qualitative picture of a magnetic field induced asymmetry of a neutrino-driven kick.
Only the large asymmetry fraction, implying a very large $B$ field, stretches this
scenario.

	As usual, more observations will provide improved tests of this picture.
In particular we would like to obtain more robust X-ray image measurement 
of pulsars with low space velocities. Improved polarization sweep position
angles for a larger data set will help as well. However, the real key to
progress is accurate measurements of proper motion vectors for more young
pulsars, especially slow objects. This means that we need improved sensitivity
to reach to higher distances and long baselines. High bandwidth VLBI studies 
extending over a decade are the key. To push to larger distances, we will
also want improved distance estimates to allow the $\vartheta$ measurements to
be corrected for differential Galactic rotation. The real challenge in collecting
more `3-D pulsar' measurements is the difficulty in estimating initial spin
periods. While some estimates may be obtained from SNR and PWN modeling
\citep[e.g.][]{van01}, most will require either historical remnants
or very regular shells with high precision expansion ages. There are not many such
pulsars, and measuring braking indices and precise proper motion for these
will not be easy.  However, with improved X-ray measurements
for robust orientations, improved radio measurements for polarization and motions,
and improved population modeling, we can hope for a more refined exploitation
of this new probe of neutron star birth dynamics.
\acknowledgments

	This work was supported by NASA grant NAG5-13344 and by CXO grants G05-6058 and
AR6-7003C issued by the Chandra X-ray Observatory Center, which is operated by the
Smithsonian Astrophysical Observatory for and on behalf of the National Aeronautics
Space Administration under contract NAS8-03060.

\vfill\eject
\clearpage

\clearpage

\clearpage

\clearpage

\clearpage

\clearpage

\clearpage

\clearpage

\clearpage

\clearpage

\clearpage

\clearpage

\clearpage

\clearpage

\begin{deluxetable}{lcccccc}
\tablecaption{Pulsar sample used in the analysis.\label{t1}}
\tablewidth{0pt}
%\tabletypesize{\tiny} %switch back to scriptsize in final version
\tabletypesize{\scriptsize}
\tablehead{
\colhead{Pulsar} & \colhead{$v\;\mathrm{(km\;s^{-1})}$} &
\colhead{$\Delta v\tablenotemark{a}\mathrm{(km\;s^{-1})}$} &
\colhead{$\vartheta_{\Omega \cdot v}$ (\arcdeg)} & \colhead{$P_0\;$(ms)} &
\colhead{$\zeta (\arcdeg)$} & \colhead{Ref.}
}
\startdata
\sidehead{1-D Pulsars}
B0114+58 & 190 & $\pm$63 & \nodata & \nodata & \nodata & \nodata \\
B0136+57 & 300 & $\pm$68 & \nodata & \nodata & \nodata & \nodata \\
B0355+54 & 61 & +12/-9 & \nodata & \nodata & \nodata & \nodata \\
B0402+61 & 653 & $\pm$121 & \nodata & \nodata & \nodata & \nodata \\
B0450+55 & 175 & +20/-19 & \nodata & \nodata & \nodata & \nodata \\
B0458+46 & 72 & $\pm$33 & \nodata & \nodata & \nodata & \nodata \\
B0540+23 & 215 & $\pm$78 & \nodata & \nodata & \nodata & \nodata \\
J0633+1746 & 128 & $\pm$3 & \nodata & \nodata & \nodata & \nodata \\
B0656+14 & 60 & +7/-6 & \nodata & \nodata & \nodata & \nodata \\
B0736-40 & 238 & +200/-23 & \nodata & \nodata & \nodata & \nodata \\
B0834+06 & 157 & +20/-19 & \nodata & \nodata & \nodata & \nodata \\
B1702-19 & 330 & $\pm$152 & \nodata & \nodata & \nodata & 1 \\
B1749-28 & 630 & $\pm$280 & \nodata & \nodata & \nodata & 1 \\
B1818-04 & 129 & +21/-21 & \nodata & \nodata & \nodata & \nodata \\
B1848+13 & 204 & +25/-25 & \nodata & \nodata & \nodata & \nodata \\
B2020+28 & 256 & +114/-61 & \nodata & \nodata & \nodata & \nodata \\
B2021+51 & 115 & +18/-15 & \nodata & \nodata & \nodata & \nodata \\
B2022+50 & 244 & $\pm$33 & \nodata & \nodata & \nodata & \nodata \\
B2217+47 & 337 & $\pm$74 & \nodata & \nodata & \nodata & \nodata \\
B2224+65 & 1605 & +193/-188 & \nodata & \nodata & \nodata & \nodata \\
\sidehead{2-D Pulsars}
B0628-28 & 318 & +61/-64 & $5\pm4$ & \nodata & \nodata & 2 \\
B0740-28 & 259 & +190/-149 & $7\pm5$ & \nodata & \nodata & 2 \\
B0823+26 & 189 & +55/-34 & $21\pm7$ & \nodata & \nodata & 3 \\
B0835-41 & 170 & $\pm$30 & $13\pm11$ & \nodata & \nodata & 3 \\
B0919+06 & 506 & $\pm$80 & $32\pm17$ & \nodata & \nodata & 3 \\
B1133+16 & 639 & +38/-35 & $22\pm2$ & \nodata & \nodata & 2 \\
B1325-43 & 597 & $\pm$254 & $31\pm22$ & \nodata & \nodata & 3 \\
B1426-66 & 150 & +40/-24 & $5\pm9$ & \nodata & \nodata & 2 \\
B1449-64 & 219 & +55/-18 & $1\pm3$ & \nodata & \nodata & 2 \\
B1508+55 & 1082 & +103/-90 & $23\pm7$ & \nodata & \nodata & 3, 4 \\
B1642-03 & 160 & +34/-32 & $26\pm5$ & \nodata & \nodata & 2 \\
B1800-21 & 347 & +48/-57 & $7\pm8$ & \nodata & $90\pm2$ & 6 \\
B1842+14 & 512 & +51/-50 & $5\pm15$ & \nodata & \nodata & 2 \\
B1929+10 & 173 & +4/-5 & $16\pm2$ & \nodata & \nodata & 2 \\
B2045-16 & 304 & +39/-38 & $3\pm6$ & \nodata & \nodata & 3 \\
IC 443 & 250 & $\pm$50 & $45\pm10$ & \nodata & \nodata & 5 \\
\sidehead{3-D Pulsars}
J0205+6449 & 838 & $\pm$251 & $21\pm10$ & $60\pm10$ & $91.0\pm0.2$ & 6 \\
B0531+21 & 140 & $\pm$8 & $26\pm3$ & $19\pm2$ & $61.3\pm0.1$ & 6 \\
J0537-6910 & 634 & $\pm$50 & $3\pm5$ & $12\pm4$ & $92.8\pm0.8$ & 6 \\
J0538+2817 & 407 & +116/-74 & $12\pm4$ & $139\pm2$\tablenotemark{b} & \nodata & 6 \\
B0540-69 & 1300 & $\pm$612 & $34\pm33$ & $38\pm2$ & $93.7\pm5$ & 6 \\
B0833-45 & 61 & $\pm$2 & $10\pm2$ & $30\pm20$ & $63.6\pm0.1$ & 6 \\
B1706-44 & 645 & $\pm$194 & $35\pm10$ & $70\pm10$ & $52\pm2$ & 6 \\
J1833-1034 & 125 & $\pm$30 & $16\pm15$ & $58\pm3$ & $85.4\pm0.3$ & 6 \\
B1951+32 & 273 & $\pm$11 & $18\pm5$ & $27\pm6$ & $\sim90\pm30$ & 6
\enddata
\tablenotetext{a}{The errors in velocity are compiled from either the uncertainties in proper motion measurement, or distance estimates, whatever larger.}
\tablenotetext{b}{An uncertainty of 20\,\,ms is used in the calculation, as
to represent a group of objects with slow spin.}
\tablerefs{Unless specified otherwise, all pulsar velocities are from \citet{hob05}
and references therein.
(1) \citet{zou05}; (2) \citet{joh05}; (3) \citet{wan06a}; (4) \citet{cha05};
(5) \citet{gae06};
(6) --see text.}
\end{deluxetable}

\begin{deluxetable}{lcccc}
\tablecaption{Best-fit parameters of the kick models. The best-fit values are obtained
from the lightly smoothed curves in Fig.~\ref{f5} $-$ \ref{f7}.
\label{t2}}
\tablewidth{0pt}
\tabletypesize{\footnotesize}
\tablehead{
\colhead{} & \colhead{$\bar{\eta}$} & \colhead{$\bar{\Omega}$} &
\colhead{$\tau$} & \colhead{$\sigma$}
}
\startdata
\sidehead{Quasi-stat}
Full Sample & $0.05^{+0.01}_{-0.02}$ & $ 26^{+37}_{-11}$ & $ 0.9^{+1.9}_{-0.7}$ & $ 0.39^{+0.14}_{-0.19}$ \\ 
Minimal Set & $0.05^{+0.04}_{-0.03}$ & $ 36^{+19}_{-23}$ & $ 1.4^{+3.7}_{-1.0}$ & $ 0.5^{+0.5}_{-0.2}$ \\ 
1D+2D & $0.13^{+0.03}_{-0.10}$ & $ 30^{+14}_{-10}$ & $ 4^{+2}_{-3}$ & $ 0.6^{+0.6}_{-0.17}$ \\ 
1D & $0.24^{+0.03}_{-0.21}$ & $ 13^{+9}_{-3}$ & $ 5^{+3}_{-5}$ & $ 0.5^{+0.6}_{-0.15}$ \\
\sidehead{Static Model}
Full Sample & $0.07^{+0.03}_{-0.02}$ & $ 9^{+10}_{-4}$ & $ 1.0^{+2.1}_{-0.9}$ & $ 0.32^{+0.07}_{-0.18}$ \\ 
Minimal Set & $0.07^{+0.04}_{-0.03}$ & $ 9^{+11}_{-3}$ & $ 0.8^{+2.1}_{-0.73}$ & $ 0.32^{+0.28}_{-0.14}$ \\ 
1D+2D & $0.22^{+0.04}_{-0.12}$ & $ 14^{+6}_{-3}$ & $ 1.4^{+0.6}_{-1.2}$ & $ 0.6^{+0.6}_{-0.17}$ \\ 
1D & $0.26^{+0.20}_{-0.17}$ & $ 7^{+3}_{-5}$ & $ 2.3^{+1.9}_{-2.1}$ & $ 0.4^{+0.5}_{-0.26}$ \\
\sidehead{$\nu$-transport}
Full Sample & $0.13^{+0.02}_{-0.06}$ & $ 7^{+1}_{-3}$ & $ 3.3^{+0.4}_{-1.5}$ & $ 0.12^{+0.03}_{-0.03}$ \\ 
Minimal Set & $0.13^{+0.04}_{-0.08}$ & $ 8^{+20}_{-3}$ & $ 3.3^{+0.5}_{-2.5}$ & $ 0.14^{+0.23}_{-0.03}$ \\ 
1D+2D & $0.13^{+0.10}_{-0.06}$ & $ 7^{+27}_{-3}$ & $ 3.5^{+0.2}_{-2.5}$ & $ 0.13^{+1.2}_{-0.03}$ \\ 
1D & $0.17^{+0.16}_{-0.05}$ & $ 8^{+5}_{-5}$ & $ 1.4^{+1.9}_{-0.5}$ & $ 0.19^{+0.5}_{-0.05}$
\enddata
\end{deluxetable}
\end{document}